\documentclass[10pt]{elsarticle}
 
\usepackage{hyperref}
\usepackage{float}
\usepackage{mathdots}

\usepackage{amscd,amssymb,amsmath,latexsym,bm}
\usepackage[mathcal,mathscr]{euscript}
\usepackage{lipsum}
\usepackage{amsfonts}
\usepackage{graphicx}
\usepackage{epstopdf}
\usepackage{algorithmic}
\usepackage{hyperref}
\usepackage{xcolor}
\usepackage{upgreek}
\usepackage{enumerate}
\usepackage{ulem}
\usepackage{gensymb}			
\usepackage[utf8]{inputenc}
\usepackage{tikz}
\usepackage{braids}
 
\newtheorem{theorem}{Theorem}[section]

\newtheorem{remark}[theorem]{Remark}

\numberwithin{equation}{section}
\numberwithin{figure}{section}

\newcommand{\CM}{{\mathbb C}}

\newcommand{\RM}{{\mathbb R}}

\newcommand{\TM}{{\mathbb T}}
\newcommand{\ZM}{{\mathbb Z}}

\newcommand{\UM}{{\mathbb U}}

\newcommand{\Bb}{{\mathcal B}}

\newcommand{\Vv}{{\mathcal V}}

\newcommand{\Tt}{{\mathcal T}}

\newcommand{\Kk}{{\mathcal K}}
\newcommand{\Hh}{{\mathcal H}}

\begin{document}

	\title{Braiding Flux-Tubes in Topological Quantum and Classical Lattice Models from Class-D}

	\author{Yifei Liu, Yingkai Liu}
	
	\address{School of Physics, Nankai University, Tianjin 300071, China}
	
 \author{Emil Prodan}
	
\address{Department of Physics, Yeshiva University, New York, NY 10016, USA}
	
\begin{abstract} 
We use magnetic flux-tubes to stabilize zero-energy modes in a lattice realization of a 2-dimensional superconductor from class D of classification table of topological condensed matter systems. The zero modes are exchanged by slowly displacing the flux-tubes and an application of the adiabatic theorem demonstrates the geometric nature of the resulting unitary time-evolution operators. Furthermore, an explicit numerical evaluation reveals that the evolutions are in fact topological, hence supplying a representation of the braid group, which turns out to be non-abelian. This physical representation is further formalized using single-strand planar diagrams. Lastly, we discuss how these predictions can be implemented with and observed in classical meta-materials and how the standard Majorana representation of the braid group can be generated by measuring derived physical observables.
\end{abstract}

\date{\today}

	\maketitle
	
	\setcounter{tocdepth}{1}

{\scriptsize \tableofcontents}

\section{Introduction}

In strongly correlated condensed matter systems, fractional  and non-abelian statistics of topological defects, called anyons, have been predicted and simulated in a large number of models \cite{BarabanPRL2009,ProdanPRB2009,
KapitPLR2012,WuPRL2014}. These properties have been recognized for their potential in quantum computing, because controlled braiding of anyons induce unitary transformations of their Hilbert sub-space, which can be concatenated into quantum gates \cite{BonesteelPRL2005}. Since the braids are topological in nature, these unitary transformations do not depend on the geometric details of the braids, hence they can be physically reproduced again and again with high fidelity. Furthermore, the anyons do not couple directly with the fermionic degrees of freedom, hence they are immune to the environment fluctuations and this, together with the topological nature of the braids, supply an error protection mechanism at the hardware level \cite{KitaevBook,FreedmanCMP2002,KitaevAOP2003,WangBook,
PachosBook}.

\vspace{0.2cm}

Evidence of non-abelian statistics in correlated electron systems is experimentally sought, indirectly,  from anyon interferometry \cite{ChamonPRB1997,BondersonAPhys2008,
WillettRPP2013},  topological spectroscopies \cite{ChangRMP2003,CooperPRL2015,MorampudiPRL2017,
PapicRPX2018}. However, demonstrating controlled concatenated braidings of the anyons, {\it e.g.} of Majorana fermions via the T-junction procedure \cite{AliceaNatPhys2011}, is quite far in the future. In a recent work \cite{BarlasArxiv2019}, one of the authors pointed out that topological point defects can be stabilized and braided in classical meta-materials. This offers new straightforward experimental venues where controlled braiding and direct observation of non-abelian statistics can be achieved. Let us point out that braiding of four Majorana-like modes in an experimental photonic setup was recently achieved in \cite{XuSciAdv2018}.

\vspace{0.2cm}

Following a similar strategy as in \cite{BarlasArxiv2019}, we here present a 2-dimensional lattice model where large numbers of topological point defects can be stabilized by magnetic flux-tubes, to form a highly quasi-degenerate resonant level in the middle of a topological spectral gap. These are somewhat similar to the Aharonov-Casher zero modes \cite{AharonovPRA1979}, whose braiding characteristics have been analyzed in \cite{KennethAP2014}. There are, however, several important differences. Our model is on a discrete lattice rather than on the continuum plane and we insisted on this feature because such models can be more straightforwardly implement experimentally with meta-materials (see section \ref{Sec:ClassicalD}). Also, in our case, the spectral stabilization of the defects modes in the middle of the topological gap is due to a particle-hole symmetry of the model, which is achieved only at flux values of $\pi$ in the natural units. 

\vspace{0.2cm}

Since the models are formulated on a lattice, there is a fundamental difficulty related to displacing the flux-tubes, because the Peierls factors behave discontinuosly when links are crossed. Nevertheless, by properly augmenting the hopping amplitudes, we succeeded on adiabatically and smoothly exchanging the flux-tubes and on ultimately obtaining an adiabatic representation of the braid group. This representation is numerically computed following the procedure from \cite{ProdanPRB2009} and the calculations confirm that the representation is non-abelian. The physical representation of the braid group is further analyzed in terms of single-strand planar diagrams for arbitrary number of flux tubes.

\vspace{0.2cm}

As demonstrated in \cite{BarlasPRB2018}, any quantum lattice model can be simulated with classical waves supported by passive meta-materials, in a manner that respects all the symmetries, in particular, the anti-unitary particle-hole symmetry. Using the algorithm discovered in \cite{BarlasPRB2018}, we supply concrete tight-binding dynamical matrices that can be implemented with networks of coupled mechanical resonators, such as springs and balls \cite{ProdanNatComm2017} or magnetically coupled spinners \cite{ApigoPRM2018}. As we shall see, implementing the braid operations in a laboratory can be achieved by slowly changing the coupling strengths between the resonators and no transfers of mass or re-configurations of the system is required.

\vspace{0.2cm}

The representations of the braid group generated by such adiabatic deformations are fundamentally different from the one generated by braiding Majorana fermions in a superconductor. The latter, however, is an induced representation which simply lifts the braid group representations  generated at BdG effective level to the many particle realm. Guided by this fact, we introduce certain derived physical observables that reside in the Clifford algebra associated to the classical zero-mode space and demonstrate that the derived representation coincides with the one generated by brading the Majorana fermions.
	
\section{The Bulk Model}

In this section we introduce the translational invariant bulk model defined on a 2-dimensional lattice and we discuss its spectral and topological characteristics as well as its particle-hole symmetry.

\subsection{The Bulk Hamiltonian and its Spectral Properties}

\vspace{0.2cm}

We will work with the minimal 2-dimensional model from the D-class of classification table of topological condensed matter \cite{SRFL2008,QiPRB2008,Kit2009,RSFL2010}, which relates to BdG effective Hamiltonians describing the fermionic excitations in a $p+\imath p$ super-conductor. The model is formulated on a square lattice and it has two degrees of freedom per each node. Hence, the Hilbert space is $\CM^2 \otimes \ell^2(\ZM^2)$. If $S_j$'s are shift operators on the lattice:
	\begin{equation}
	S_j |\bm m \rangle =| m + \bm e_j \rangle, \quad S_j^\dagger |\bm m \rangle = |\bm m - \bm e_j \rangle, \quad j=1,2,
\end{equation}
where $\bm e_j$ are the generators of $\ZM^2$, then the Hamiltonian of our model can be written in the following compact form:
\begin{equation}\label{Eq:BulkH}
	H=\sigma_3\otimes\bigg [ M \, I+\tfrac{1}{2}  \sum_{j=1,2} (S_j +S_j^\ast)\bigg ] + \tfrac{1}{2 \imath} \sum_{j=1,2} \sigma_j \otimes (S_j -S_j^\ast).
\end{equation}
Throughout, the $\sigma$'s will denote Pauli's matrices. As one can see, we fixed the energy and space units such that the Hamiltonian displays only one parameter, the mass $M$. 

\vspace{0.2cm}

\begin{figure}[t]
\centering
\includegraphics[width=\textwidth]{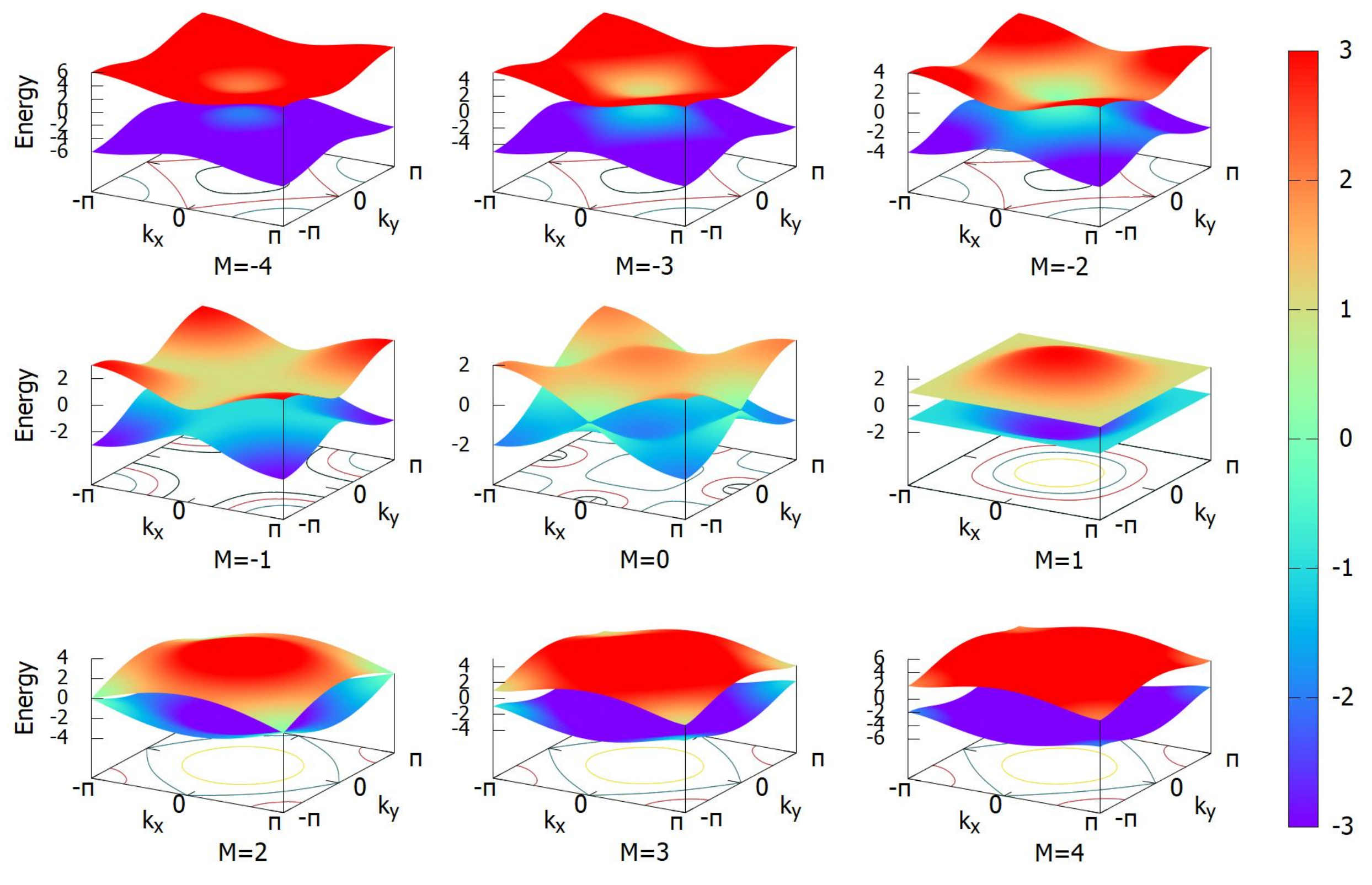}
\caption{Band structure of the Hamiltonian \eqref{Eq:BulkH} for selected values of $M$ parameter. Band touchings are observed at $M= \pm 2$ and $M=0$.}
\label{Fig:EnergyKxKy}
\end{figure}

The energy dispersion spectrum can be computed explicitly:
\begin{equation}
E_{\bm k} = \sqrt{ \Big(M+\sum_{j=1,2}\cos(k_j)\Big)^2+\sum_{j=1,2} \sin(k_j)^2 }, \quad \bm k \in [-\pi,\pi]^2,
\end{equation} 
and plots of the dispersion bands are supplied in Fig.~\ref{Fig:EnergyKxKy}. The data reveals bulk gap closings at $M=0$ and $\pm 2$ and, as we shall see, these are all topological phase transitions.

\subsection{Topological characteristics}

\vspace{0.2cm}

\begin{figure}[t]
\centering
\includegraphics[width=\textwidth]{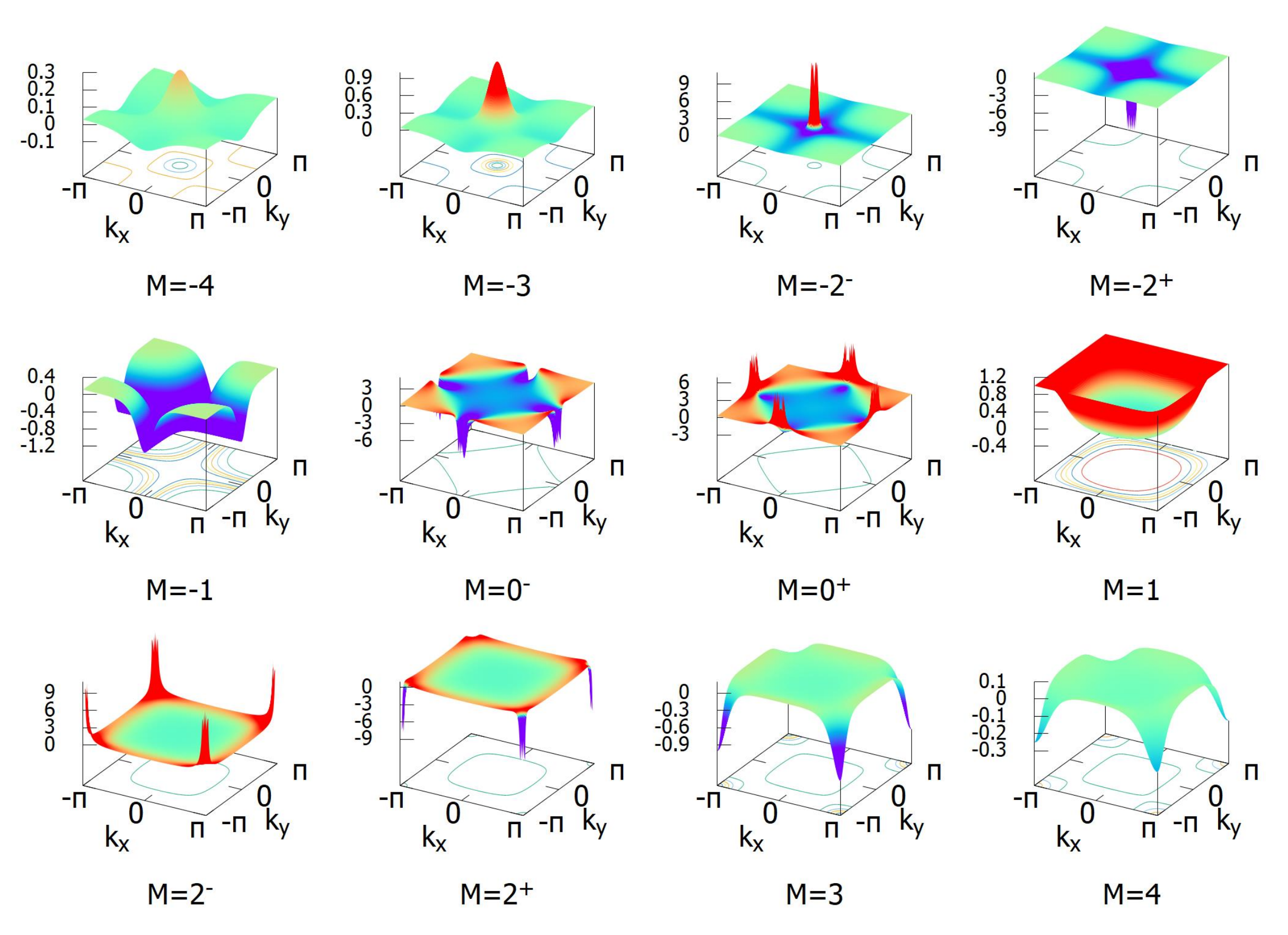}
\caption{Plots of the Berry curvature as function of quasi-momentum, for selected values of $M$ parameter.}
\label{Fig:BerryCurvature}
\end{figure}

 The Berry curvature of the lower band can be also computed explicitly: 
	\begin{align}
	F(\bm k) =\frac{\cos k_1+\cos k_2+M \cos k_1 \cos k_2}{\big ((M+\cos k_1+\cos k_2)^2+\sin k_1 ^2+\sin k_2^2\big)^{3/2}}	
	\end{align}
and plots of it are supplied in Fig.~\ref{Fig:BerryCurvature}. The bulk invariant, which labels the topological phases of the model, is the Chern number of the lower band:
	\begin{align*}
	\int_{\TM^2}{\rm d} \bm k \ F(\bm k)=
	\begin{cases}
	1&\text{for}\ 0<M<2\\
	-1&\text{for}\ -2<M<0\\
	0&\text{for}\ 2<|M|.
	\end{cases}
	\end{align*}
There are two topological phases with a common boundary at $M=0$ and these topological phases are surrounded by a trivial phase which resides in the region $|M|>2$ of the parameter space.

\vspace{0.2cm} 

The hallmark of the topological phases is the emergence of chiral edge states whenever a bulk sample is halved \cite{ProdanBook1}. In this work, however, we will be interested in point defects rather than edges. Stabilization of zero-dimensional topological modes inside the bulk gap is enabled by an additional and fundamental property of the model, namely, its particle-hole symmetry:
\begin{equation}\label{Eq:PHSymmetry}
\Theta_{\rm PH} H \Theta_{\rm PH}^{-1} = -H, \quad \Theta_{\rm PH} = (\sigma_1 \otimes I) \mathcal K, \quad \Theta_{\rm PH}^2 = 1,
\end{equation} 
where $\Kk$ is the ordinary complex conjugation. This symmetry forces the energy spectrum to be symmetric relative to $E=0$ mark.
	
\section{Magnetic Flux Insertion}

In this section, we insert infinitely thin magnetic flux-tubes through the plane of the lattice and  study the spectral properties of the resulting Hamiltonian. As a result of the topological character of the Hamiltonian and of its particle-hole symmetry, a quasi-degenerate energy level develops at zero energy when the flux values are properly adjusted. The invariant space corresponding to this quasi-degenerate level will supply the representation space for the braid group.

\subsection{Peierls substitution} 

\vspace{0.2cm}

\begin{figure}[t]
\centering
\includegraphics[width=0.7\textwidth]{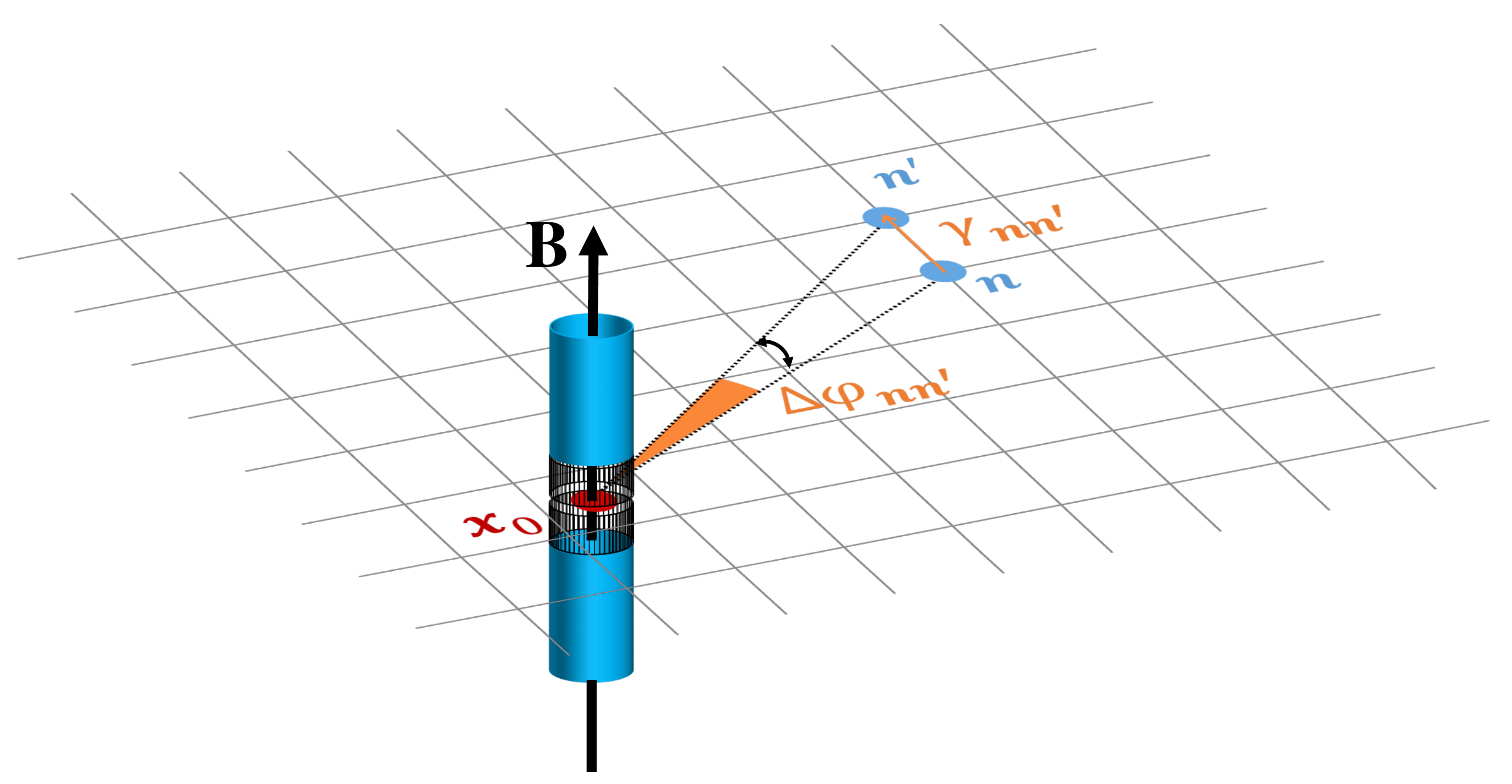}
\caption{Geometry associated with the flux-tube insertion.}
\label{Fig:FluxInsertion}
\end{figure}

For a generic lattice Hamiltonian on the Hilbert space $\CM^2 \otimes \ell^2(\ZM^2)$:
	\begin{align}
	H=\sum_{\bm n,\bm n'} h_{\bm n',\bm n} \otimes |\bm n ' \rangle \langle \bm n |,
	\end{align} 
a flux-tube insertion at position $\bm x$ is taken into account by the Peierls substitution \cite{Peierls}, which transforms the Hamiltonian into: 
	\begin{align}
	H(\bm x)=\sum_{\bm n,\bm n'} \exp \Big (\imath \int_{\gamma_{\bm n \bm n'}}\vec{A}\cdot{\rm d} \vec{\ell} \, \Big) \, h_{\bm n',\bm n} \otimes |\bm n'\rangle \langle\bm n |,
	\end{align} 
	where $\vec{A}$ is the corresponding magnetic vector potential. A geometric representation of the situation is supplied in Fig.~\ref{Fig:FluxInsertion}.	For a flux-tube of infinitesimal radius and carrying a magnetic flux $\Phi$, the vector potential takes the form $\vec A = \frac{\Phi}{2 \pi r} \hat e_\varphi$ in the standard polar coordinates centered at $\bm x$. As such, the Peierls phase has an explicit expression:
	\begin{equation}
\theta_{\bm n\bm n'}(\bm x) =	\int_{\gamma_{\bm n \bm n'}}\vec{A}\cdot {\rm d}{\vec{\ell}}= \frac{\Phi}{2\pi} \Delta \varphi_{\bm n\bm n'},
	\end{equation}
with the $\Delta \varphi$ angles as specified in Fig.~\ref{Fig:FluxInsertion}. 

\begin{remark}\label{Re:Peierls}{\rm Several important observations are in place:
\begin{itemize}
\item The angles $\Delta \varphi_{\bm n \bm n'}$ are measured from $\bm n$ towards $\bm n'$, hence they come with definite sign, which is positive if the rotation of $\bm n$ towards $\bm n'$ is in the trigonometric sense, and negative otherwise.
\item Up to an additive factor of $\pm 2\pi$, $\Delta \varphi_{\bm n \bm n'}=\varphi_{\bm n'}-\varphi_{\bm n}$, where the latter are the angular coordinates of $\bm n$ and $\bm n'$, respectively, in a coordinate system centered at $\bm x$. This statement can be written more precisely as:
\begin{equation}\label{Eq:Mod2PI}
\Delta \varphi_{\bm n \bm n'}=\big (\varphi_{\bm n'}-\varphi_{\bm n}\big ) {\rm mod}\, 2\pi.
\end{equation} 
\item All angles defined above depend on the position of the flux-tube and, when needed, we will specify this dependency explicitly. In particular, $\Delta \varphi_{\bm n \bm n'}(\bm x)$ jumps by $\pm 2\pi$ whenever $\bm x$ crosses the segment joining $\bm n$ and $\bm n'$. This will have  important consequences for the braiding of the flux-tubes.
\end{itemize}
}
\end{remark}

With the above data, our specific model with $N$ flux-tubes inserted at positions $\bm x_1,\ldots,\bm x_N$ becomes:
\begin{align}\label{Eq:Hamiltonian}
& H(\bm x_1|\Phi_1,\ldots,\bm x_N|\Phi_N)=  M \sigma_3 \otimes I \\ \nonumber 
& \quad +\tfrac{1}{2} \sum_{j=1,2}\sum_{\bm n \in \mathbb Z^2} \bigg [ e^{\imath \theta_{\bm n, \bm n+\bm e_j}}  (\sigma_3 - \imath \sigma_j)\otimes |\bm n +\bm e_j\rangle \langle \bm n|+ h.c. \bigg] ,
\end{align}
with $\theta_{\bm n \bm n'}=\sum_{k=1}^N \theta_{\bm n \bm n'}(\bm x_k)$. The mass parameter $M$ will be fixed in the middle of the topological phase from now on, namely, $M=1$.

\subsection{Spectral flow with the flux}

\vspace{0.2cm}
	
\begin{figure}[t]
		\centering
		\includegraphics[width=0.7\textwidth]{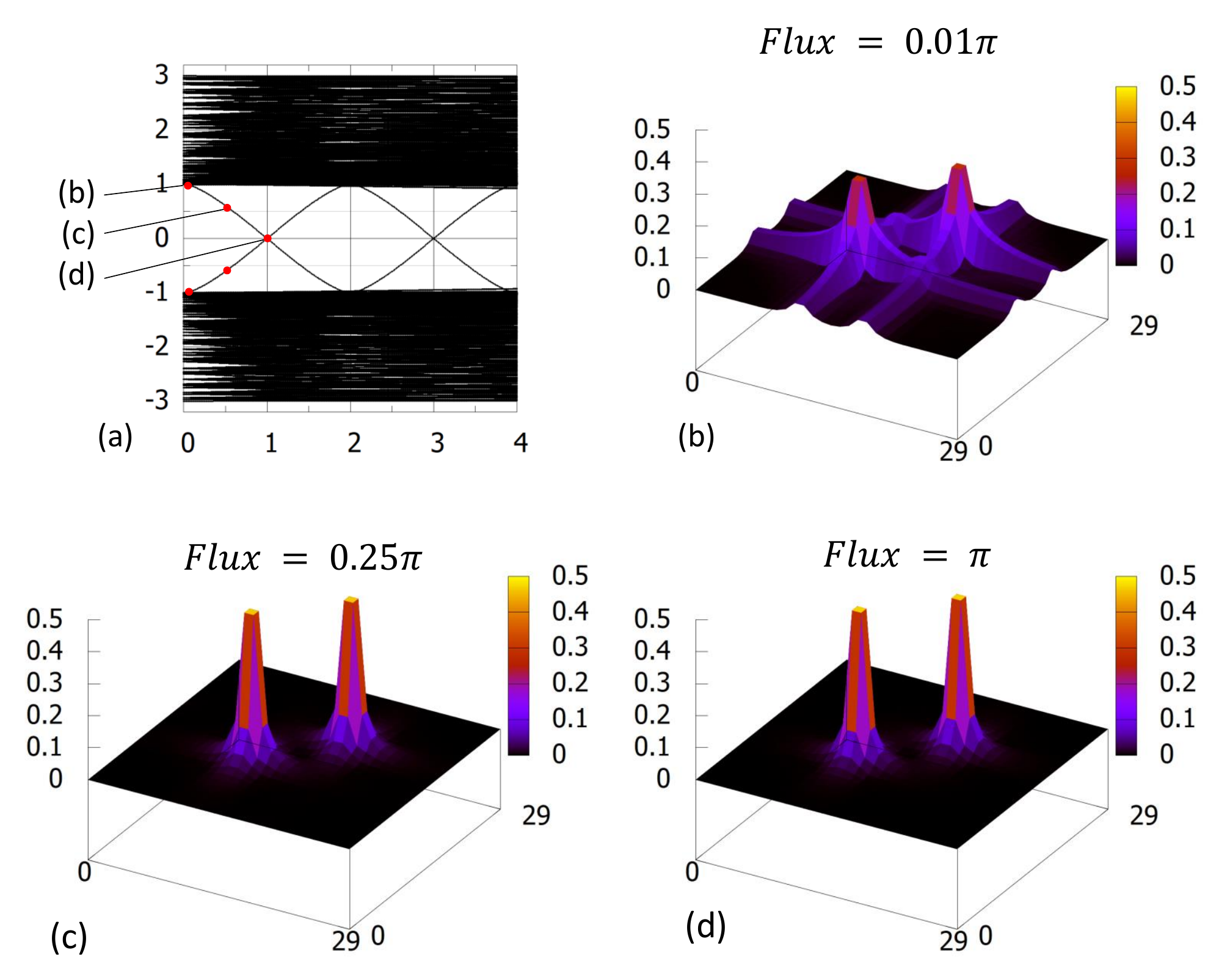}
		\caption{(a) Spectrum of $H(\bm x_1|\Phi,\bm x_2|-\Phi)$ as function of $\Phi$. (b-c) The spatial profile of the eigen-vectors associated with the eigenvalues marked by the red dots in panel (a). Spatial localization of the eigen-vectors becomes sharper and sharper as the eigenvalues approach the mid-gap level.}
	\label{Fig:SpecVsFlux}
\end{figure}

We present first numerical results for two flux-tubes positioned at fixed locations $\bm x_1$ and $\bm x_2$. Since our computations are performed with periodic boundary conditions, we are constrained to consider $\Phi_1=-\Phi_2=\Phi$, for a total of zero-flux through the entire lattice.	The spectrum of the Hamiltonian as function of $\Phi$ is reported in Fig.~\ref{Fig:SpecVsFlux}, together with samples of eigen-functions profiles. As it is well known \cite{AvronCMP1994}, the flux-tubes trap a number of electron states equal to the Chern number of the bulk model and, as the value of the flux increases/decreases, the eigen-energy of the trapped states flows from the upper/lower band towards the lower/upper band, setting in motion a spectral flow. This is precisely what can be observed in Fig.~\ref{Fig:SpecVsFlux}. Let us point out that the pair of chiral bands seen there correspond to the two flux-tubes and, since the tubes are spatially separated, there is practically no interference between  the two and they can be analyzed one at a time. In particular, the pairs of eigen-functions shown in Figs.~\ref{Fig:SpecVsFlux} are practically the topological modes trapped by the individual flux-tubes.

\vspace{0.2cm}

The spectral flow repeats itself with a period of $\Phi=2\pi$ in Fig.~\ref{Fig:SpecVsFlux}. To understand this feature in the most general context, let us go back to the general case of $N$ tubes and consider an increase by $2 \pi$ of one flux $\Phi_j$. We define the unitary transformation: 
\begin{equation}
|\bm n \rangle \rightarrow U_k(\bm x_j) |\bm n \rangle =e^{\imath k \varphi_{\bm n}(\bm x_j)} \, |\bm n \rangle, \quad \bm n \in \ZM^2, \quad k \in \mathbb Z,
\end{equation}
with the action on the hopping operators:
\begin{equation}\label{Eq:UFactor}
U_k(\bm x_j)\, |\bm n' \rangle \langle \bm n | \, U_k^{-1}(\bm x_j)= e^{\imath  k (\varphi_{\bm n'}(\bm x_j)-\varphi_{\bm n}(\bm x_j))} \, |\bm n' \rangle \langle \bm n |.
\end{equation}
When the flux $\Phi_j$ is increased by $2\pi$, the Peierls phase factor is modified by the multiplicative term $e^{\imath \Delta \varphi_{\bm n\bm n'}(\bm x_j)}$ and, from \eqref{Eq:Mod2PI}, we see that this factor is the same as the one in \eqref{Eq:UFactor} for $k=1$. Therefore:
\begin{equation}
H(\Phi_j+2\pi) = U_1(\bm x_j)H(\Phi_j)U_1^{-1}(\bm x_j),
\end{equation} 
and this explains why an increase of any of the fluxes by multiples of $2\pi$ leaves the energy spectrum unchanged. 

\subsection{The space of zero-modes}

\vspace{0.2cm}

\begin{figure}[t]
		\centering
		\includegraphics[width=0.9\textwidth]{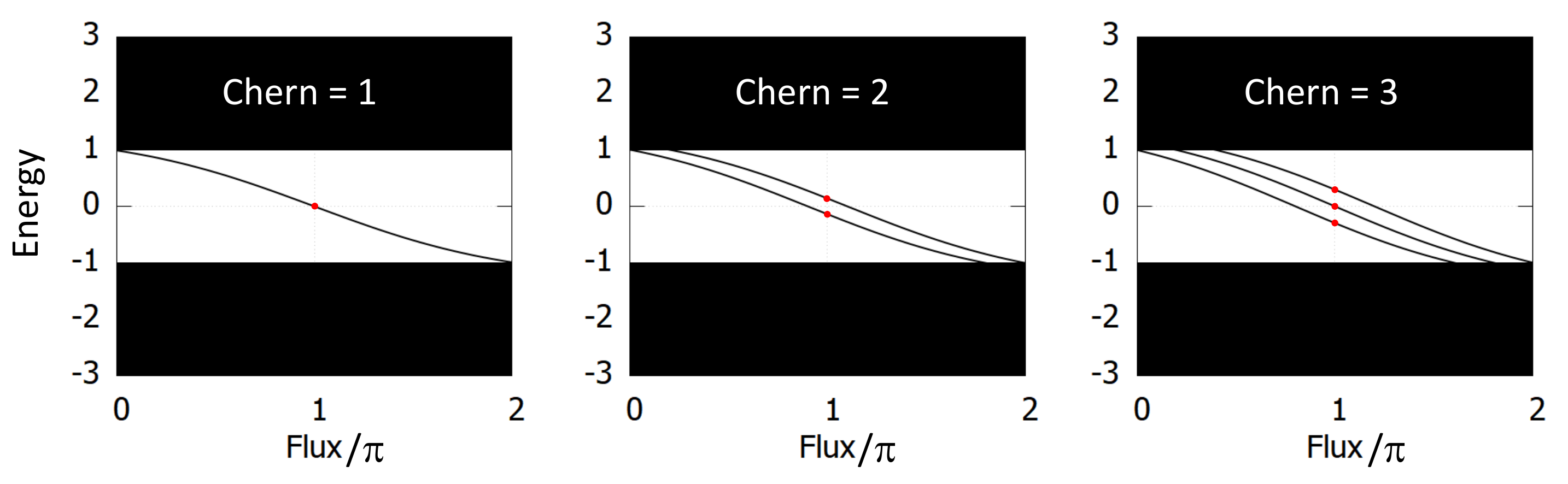}
		\caption{Spectral flow under  a single flux-tube insertion for three values of the Chern number. The models with higher Chern numbers were obtained by stacking appropriate numbers of minimal models and by turning on a soft coupling between the stacked copies. The cases with odd Chern numbers display exact zero-modes while, in the case of even Chern number, the mid-gap modes are unstable and split from $E=0$ mark.}
	\label{Fig:OddVsEvenChern}
\end{figure}

The second interesting feature in Fig.~\ref{Fig:SpecVsFlux} is the intersection of the spectral flows, which occurs at zero energy and at $\Phi = (2p +1)\pi$, $p \in \mathbb Z$. This phenomenon is related to the particle-hole symmetry of the original model. Indeed, let us consider the general case of $N$ flux tubes with $\Phi_j = (2 p_j +1)\pi$. Then a conjugation by $\Theta_{\rm PH}$ transforms the Peierls phase factors from $e^{\imath (p_j +\frac{1}{2}) \Delta \varphi_{\bm n \bm n'}(\bm x_j)}$ to $e^{-\imath (p_j +\frac{1}{2}) \Delta \varphi_{\bm n \bm n'}(\bm x_j)}$. These two terms differ by a multiplicative factor of $e^{-\imath (2p_j +1) \Delta \varphi_{\bm n \bm n'}(\bm x_j)}$ and, by recalling \eqref{Eq:Mod2PI}, this can be easily assessed to be a purely gauge conjugation. More precisely:   
\begin{align}
\Theta_{\rm PH} \, H(\bm x_1|\Phi_1,\ldots,\bm x_N|\Phi_N ) \, \Theta_{\rm PH}^{-1} = - U H(\bm x_1|\Phi_1,\ldots,\bm x_N|\Phi_N) U^{-1},
\end{align}
with:
\begin{align}\label{Eq:PHGauge}
U= U_{2p_1+1}(\bm x_1) \ldots U_{2p_N+1}(\bm x_N).
\end{align}
It is convenient to re-define the PH-symmetry operation as:
\begin{equation}
\Theta_{\rm PH} = U^{-1} (\sigma_1 \otimes I) \mathcal K, \quad \Theta_{\rm PH}^2 = I,
\end{equation}
such that:
\begin{align}
\Theta_{\rm PH} \, H(\bm x_1|\Phi_1,\ldots,\bm x_N|\Phi_N ) \, \Theta_{\rm PH}^{-1} = - H(\bm x_1|\Phi_1,\ldots,\bm x_N|\Phi_N).
\end{align}
It now becomes clear that the energy spectra must be mirror symmetric relative to $E=0$, whenever the fluxes $\Phi_j$ are odd multiples of $\pi$. This implies that, if the bulk Chern number is odd and the flux-tubes are well separated such that they can be treated separately, then each of their chiral spectral flows must cross the level $E=0$ at $\Phi=(2p+1)\pi$. This is illustrated by an explicit numerical simulation in Fig.~\ref{Fig:OddVsEvenChern}.

\vspace{0.2cm}

The conclusion is that, if $\Phi_j$'s are fixed at $\pm \pi$, the flux-tubes are well separated and our model is in a topological phase, then $H(\bm x_1|\Phi_1,\ldots,\bm x_N|\Phi_N )$ displays a $N$-fold quasi-degenerate energy level centered at zero. The spectrum corresponding to this quasi-degenerate level will be denoted by $\Sigma_0$ and eigen-vectors associated with $\Sigma_0$ will be referred as zero-modes. The linear $N$-dimensional space spanned by them will be referred as the zero-modes space and will be denoted by $\Hh_0(\bm x_1|\Phi_1,\ldots,\bm x_N|\Phi_N )$, with the understanding that $\Phi_j$'s are fixed at $\pm \pi$. Note that this sub-space of $\CM^2 \otimes \ell^2(\ZM^2)$ changes with the positions of the flux-tubes.

\section{Adiabatic Displacement of the Flux-Tubes}

In this section, we investigate the time evolution of the quantum states induced by a slow displacement of the flux-tubes. In its current form, the Hamiltonian evolves discontinuously whenever a flux-tube crosses an edge of the lattice. Our main goal here is to describe a solution which resolves this issue and, consequently, it allows us to achieve the adiabatic regime where the time evolution of the zero-modes becomes purely geometric in nature.

\subsection{Resolving the discontinuity of the Hamiltonian}
\label{SubS:ResDisc}

\vspace{0.2cm}

Let us recall the discussion from Remark~\ref{Re:Peierls}, where we learned that $\Delta \varphi_{\bm n \bm n'}$ jumps by an additive factor of $\pm 2 \pi$ whenever a flux-tube crosses the segment joining $\bm n$ and $\bm n'$. Because the flux-tubes are fixed at odd multiples of $\pi$, this implies that the Peierls phase factors in the Hamiltonian~\ref{Eq:Hamiltonian} change sign. The only way to correct for this undesirable effect, is to smoothly turn off the coefficient of the Hamiltonian that is affected by this phenomenon. This can be done by modifying the Hamiltonian to the following expression: 
\begin{align}\label{Eq:NewHamiltonian}
& \qquad \qquad H(\bm x_1|\Phi_1,\ldots,\bm x_N|\Phi_N)=  M \sigma_3 \otimes I \\ \nonumber 
& \quad +\tfrac{1}{2} \sum_{j=1,2}\sum_{\bm n \in \mathbb Z^2} \bigg [ g(\bm n,\bm n+\bm e_j)e^{\imath \theta_{\bm n, \bm n+\bm e_j}}  (\sigma_3 - \imath \sigma_j)\otimes |\bm n +\bm e_j\rangle \langle \bm n|+ h.c. \bigg] ,
\end{align}
where
\begin{equation}
g(\bm n,\bm n') =\prod_{k=1}^N \Big (1-\exp\Big(-(|0.5(\bm n+\bm n')-\bm x_k|/\gamma)^2\Big)\Big ).
\end{equation}
This particular choice was made to ensure that, whenever the position $\bm x_k$ of any of the flux-tubes traverses a link $(\bm n,\bm n +\bm e_j)$ through the middle of it, the hopping term corresponding to this link, and only to this link, is smoothly turned off by the factor $g$ inserted in the Hamiltonian. Hence, from now on, we work exclusively with the Hamiltonian \eqref{Eq:NewHamiltonian} and the flux-tubes will be displaced such that they always cross the links of the lattice through the middle. The parameter $\gamma$ will be fixed at $1$ throughout.

\begin{figure}[t]
		\centering
		\includegraphics[width=0.9\textwidth]{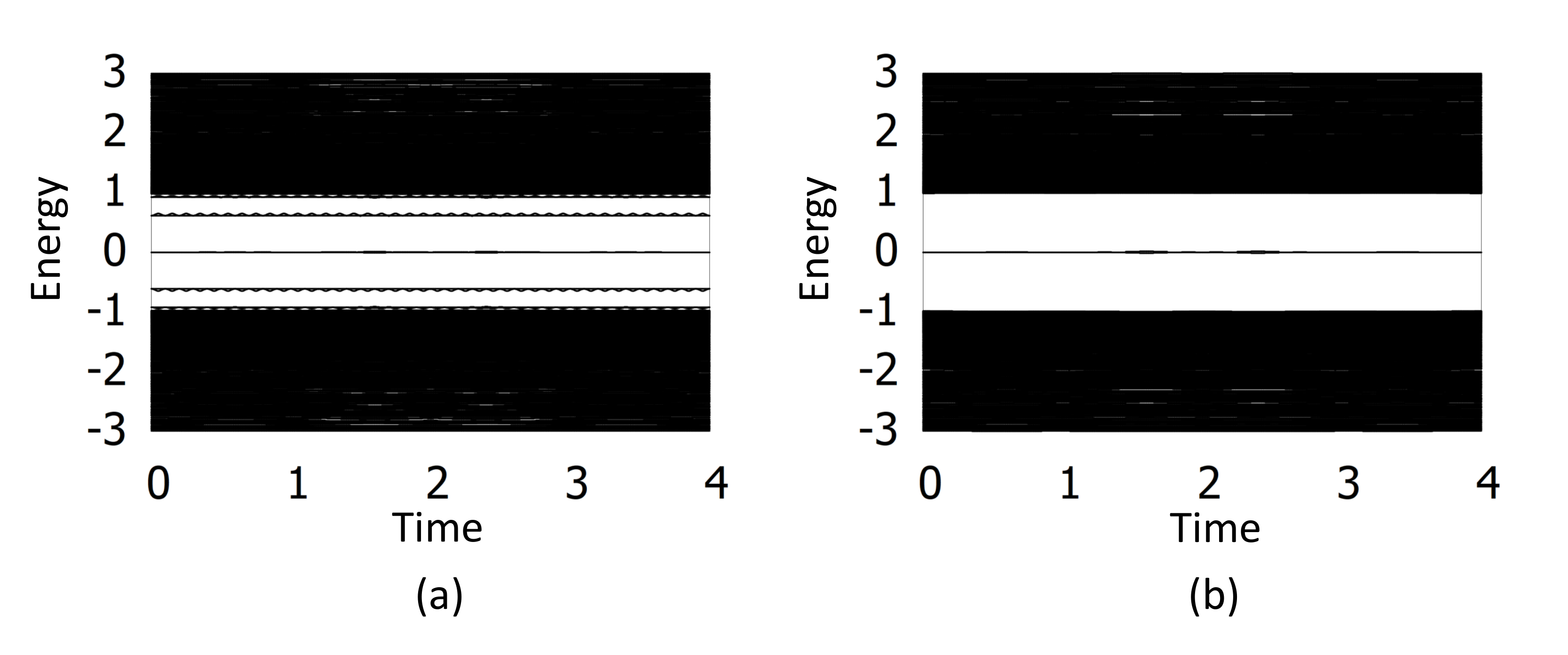}
		\caption{Evolution of the spectrum of (a) Hamiltonian~\eqref{Eq:NewHamiltonian} ($\gamma=1$), and (b) Hamiltonian~\eqref{Eq:Hamiltonian}, when a (+)-flux tube is circled around a (-)-flux tube.}
	\label{Fig:SmoothEvolution}
\end{figure}

\vspace{0.2cm}

In Fig.~\ref{Fig:SmoothEvolution}, we report the evolution of the energy spectrum computed with the Hamiltonian \ref{Eq:NewHamiltonian}, as a (+) flux-tube is circled around a (-) flux-tube. It demonstrates that the quasi-zero energy spectrum remains isolated from the rest of the spectrum, which is the pre-requisite for achieving the adiabatic regime. Further tests and comparisons between the Hamiltonians \eqref{Eq:Hamiltonian} and \eqref{Eq:NewHamiltonian} are illustrated in Fig.~\ref{Fig:UTest}, after more background is developed. As we shall see, it is practically impossible to achieve the adiabatic regime with the Hamiltonian \eqref{Eq:Hamiltonian}, due to the sudden discontinuities discussed above, but it is possible with the Hamiltonian \eqref{Eq:NewHamiltonian}.

\subsection{The time evolution operators}

\vspace{0.2cm}

In the case when the positions of the flux-tubes change in time, a quantum state $\Psi_0$ prepared at time $t_0$ evolves as:
\begin{equation}
\Psi(t)=U(t,t_0)\Psi_0,
\end{equation} 
where $U(t,t_0)$ is the unitary time evolution operator supplied by the unique solution of the equation:
\begin{equation}\label{Eq:TimeEvolution1}
\imath \partial_t U(t,t_0) = H\big (\bm x_1(t)|\Phi_1,\ldots,\bm x_N(t)|\Phi_N \big ) U(t,t_0), \quad U(t_0,t_0)=I.
\end{equation}
The following standard property of the family of evolution operators will play a role later on:
\begin{equation}\label{Eq:GroupProperty}
U(t'',t')U(t',t) = U(t'',t), \quad \forall \, t,t',t'' \in \RM.
\end{equation}

\subsection{Adiabatic time-evolution and geometric monodromies}

\vspace{0.2cm}

The time evolution considered above can be thought of as induced by the motion of a point $x$ in the multi-dimensional space $\RM^{2N}$. Consider now a path $\zeta : [0,1]\rightarrow \RM^{2N}$ in this space and a point that slides along this path with speed $v=1/T$, resulting in the time-dependent Hamiltonian:
\begin{equation}
H_t = H\big (\bm x_1(v t)|\Phi_1,\ldots,\bm x_N(vt)|\Phi_N \big ), \quad t \in [0,T].
\end{equation}
Let: 
\begin{equation}
P_x = \chi_{[-\epsilon,\epsilon]}\Big (H\big (\bm x_1|\Phi_1,\ldots,\bm x_N|\Phi_N \big ) \Big )
\end{equation}
 be the spectral projector onto the zero-modes space, when the flux-tubes are arranged in an arbitrary instantaneous configuration $x \in \RM^{2N}$. Above, $\epsilon$ is a small positive constant accounting for the spreading of $\Sigma_0$ when the flux-tubes are at finite separations from one each other and $\chi$ denotes the indicator function. This operator projects the large Hilbert space $\Hh=\CM^2 \otimes \ell^2(\ZM^2)$ onto the zero modes space:
 \begin{equation}
 P_x \Hh = \Hh_0\big (\bm x_1|\Phi_1,\ldots,\bm x_N|\Phi_N \big ).
 \end{equation}
Then the adiabatic theorem \cite{Kato1950} assures us that, under certain regularity conditions related to the smoothness of the time-dependence of $H_t$:
\begin{equation}\label{Eq:AdiabaticTh}
U(t,t_0) \,P_{x_0} =  W_\zeta (x_t)\, P_{x_0} + o(v),
\end{equation}
 where $W_\zeta(x):P_{x_0}\Hh \rightarrow P_x \Hh$ is the monodromy along $\zeta$, {\it i.e.} the unique unitary solution of the equation:
\begin{equation}
\imath \bm \nabla_x W(x) = \imath [\bm \nabla_x P_x,P_x] \, W(x), \quad W(x_0)=P_{x_0},
\end{equation}
when integrated along the path $\zeta$. Numerically, the monodromy can be conveniently computed as \cite{ProdanPRB2009}:
\begin{equation}\label{Eq:PracticalUa}
W_\zeta(x) = \lim_{K \rightarrow \infty} P_{x_K}\ldots P_{x_0},
\end{equation}
where $\{ x_0,\ldots,x_K \}$ is a  discretization of the segment of path $\zeta$ joining $x_0$ and $x$. In practice, $K$ will have to be set to a finite value which will be referred to as the discretization parameter. 

\begin{figure}[t]
\centering
\includegraphics[width=0.8\textwidth]{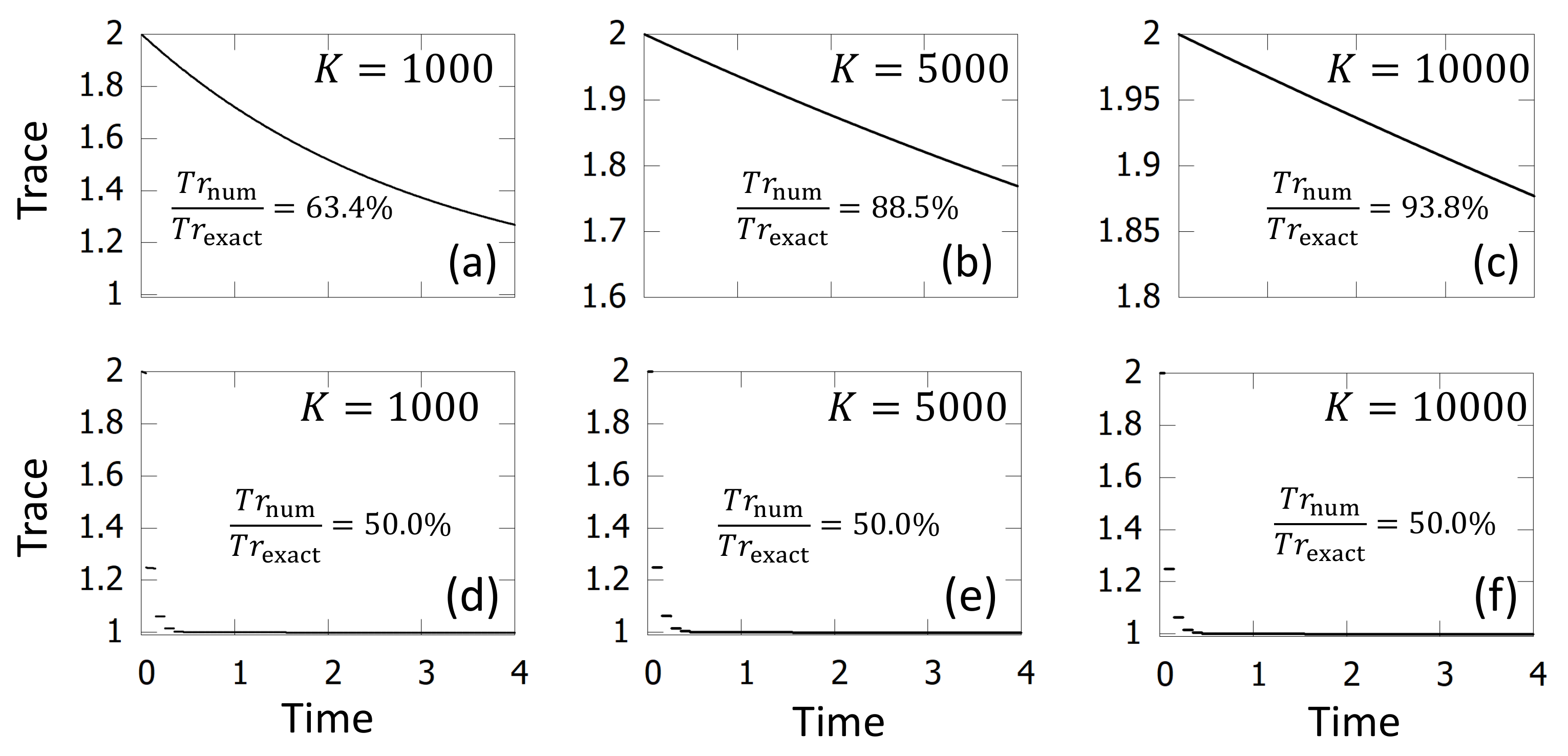}
\caption{The unitarity test \eqref{Eq:UTest2} when a (+)-flux tube is circled around a (-)-flux tube, for (a,b,c) Hamiltonian~\eqref{Eq:NewHamiltonian} ($\gamma=1$), and (d,e,f) Hamiltonian~\eqref{Eq:Hamiltonian}. The monodromies were computed with \eqref{Eq:PracticalUa} using different discretizations of the world-lines, namely $K=1000$, $5000$ and $10000$.}
\label{Fig:UTest}
\end{figure}

\vspace{0.2cm}

Note that \eqref{Eq:AdiabaticTh} is an adaptation of the general adiabatic theorem, as formulated in \cite{Kato1950}, to our particular context. It takes that form precisely because the spectrum $\Sigma_0$ remains pinned at zero for all flux-tube configurations, which is a remarkable consequence of the topological character of the model and its PH-symmetry. In these conditions, \eqref{Eq:AdiabaticTh} says that the unitary adiabatic time evolution of the zero modes is purely geometric in nature. Indeed, let $T$ be the time it takes the point $x$ to slide along $\zeta$. Then:
\begin{equation}
U(T+t_0,t_0)P_{x_0} = W_\zeta(x_f) P_{x_0} + o(1/T),
\end{equation}
and, according to the above, the time variable is completely absent in the definition of $W_\zeta$, which in fact is entirely determined by the path $\zeta$. When the curve $\zeta$ closes into itself, the unitary time evolution operator maps the initial zero modes space into itself, hence supplying a unitary operator $W_\zeta : P_{x_0} \Hh \rightarrow P_{x_0} \Hh$.

\vspace{0.2cm}

Our last comment here is that, if one moves the flux-tubes and uses the original Hamiltonian \eqref{Eq:Hamiltonian}, then one will find out that the resulting time-dependent Hamiltonian $H_t$ does not satisfy the regularity conditions required by the adiabatic theorem. It is at this point where the solution supplied in section \ref{SubS:ResDisc} is to be appreciated. To convince the reader about the striking difference between the two cases, we show in Fig.~\ref{Fig:UTest} a test on the unitarity of $W_\zeta(x)$:
\begin{equation}\label{Eq:UTest1}
W_\zeta (x)^\dagger W_\zeta (x) = I_{P_{x_0}\Hh},
\end{equation}
 as one flux-tube is circled around another, as in Fig.~\ref{Fig:SmoothEvolution}. Instead of using \eqref{Eq:UTest1} directly, we use the quantitative test:
 \begin{equation}\label{Eq:UTest2}
 {\rm Tr}\big ( W_\zeta (x)^\dagger W_\zeta (x) \big ) = {\rm dim} \, P_{x_0} \Hh.
 \end{equation}
 As one can see in Fig.~\ref{Fig:UTest}, when increasing the discretization parameter, the test becomes more and more accurate when the braiding is performed with Hamiltonian \eqref{Eq:NewHamiltonian}, but this is not at all the case when the Hamiltonian \eqref{Eq:Hamiltonian} is used instead.
 
\section{The Braid Group $\Bb_N$ and its Physical Representations}

\begin{figure}[t]
		\centering
		\includegraphics[width=0.5\linewidth]{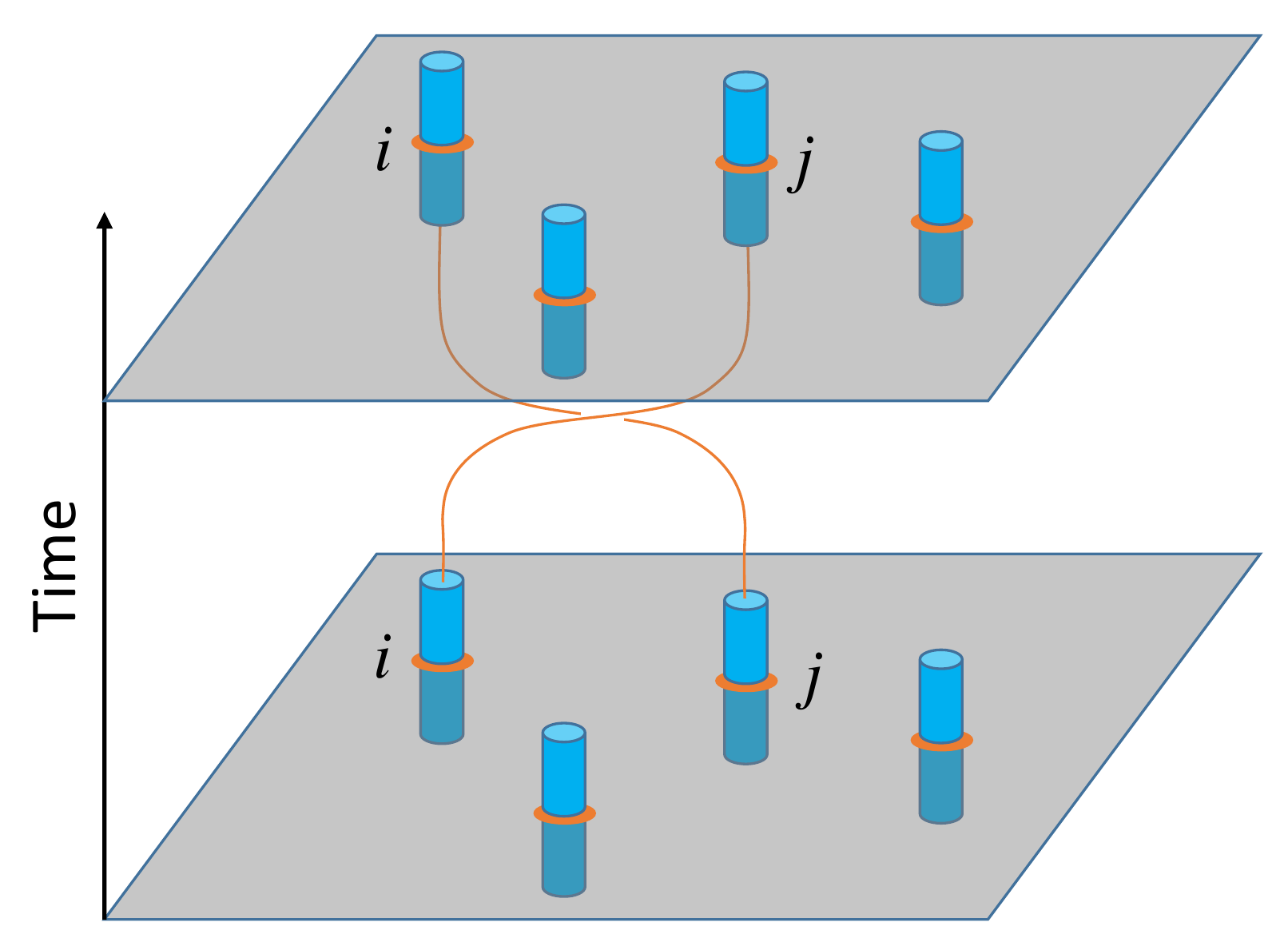}
		\caption{\small Schematic of a physical process which permutes the flux-tubes $i$ and $j$. The lines represent the world-lines of the two flux-tubes which encode their positions at any instantaneous moment of time.}
	\label{Fig:BraidingSchematic}
\end{figure}

In this section we consider physical processes as the ones illustrated in Fig.~\ref{Fig:BraidingSchematic}(a), which result in a permutation of the flux-tubes. The unitary operators induced by such actions are not fully determined by the final permutations of the flux-tubes, but they depend on the world-lines, {\it e.g.} on whether these world-lines enclose or not other flux-tubes. The goal of this section is to connect the adiabatic unitary operators resulting from such actions and the braid group \cite{BraidBook}.

\subsection{The braid group}

\begin{figure}[t]
\centering
\includegraphics[width=0.8\linewidth]{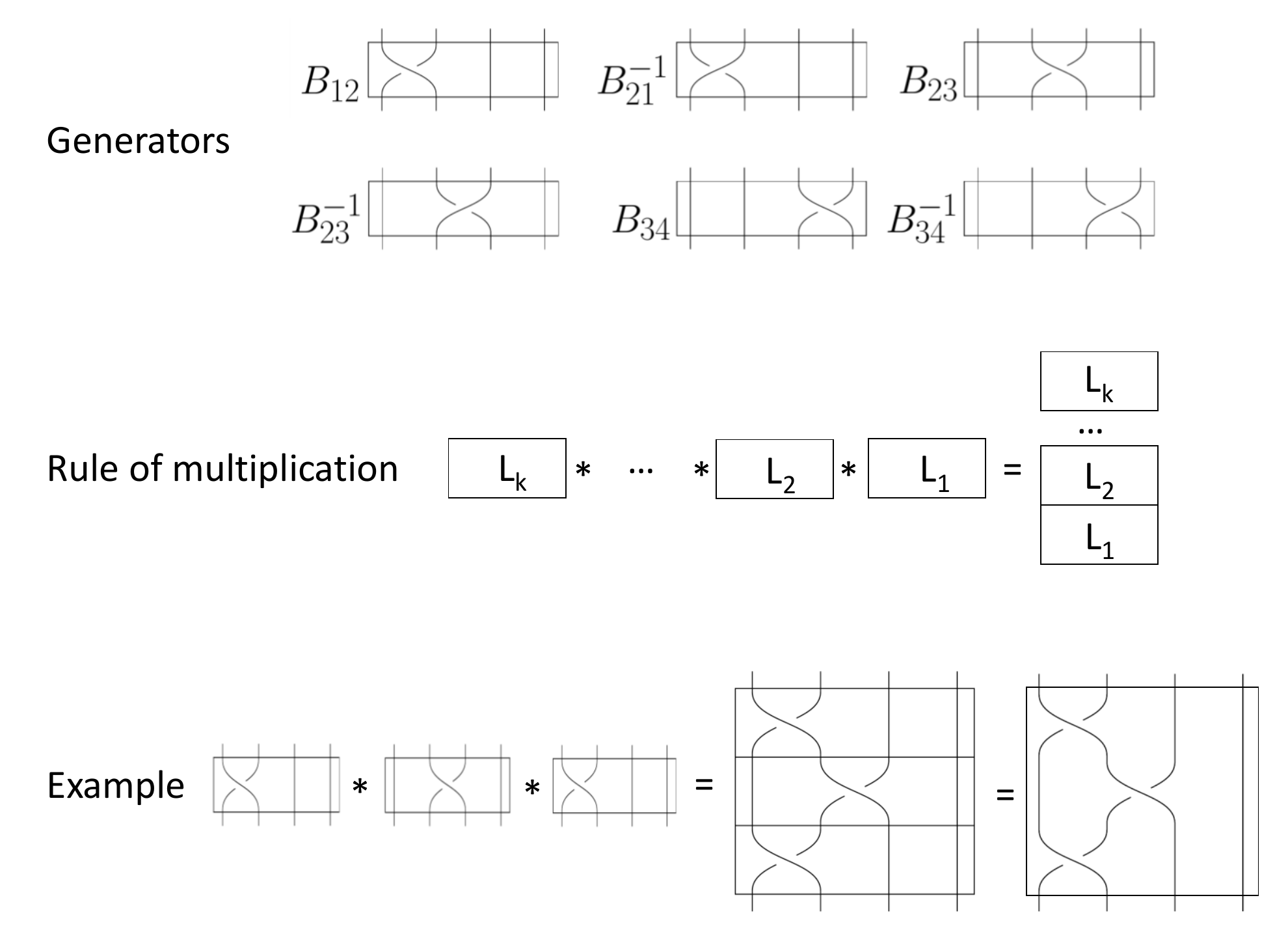}
\caption{\small {Pictorial representation of the braid group.} }
\label{Fig:PictorialBraidGroup}
\end{figure}

\vspace{0.2cm}

The braid group $\Bb_N$ consists of all finite braid operations that can be applied to $N$ strands.
Two braids can be concatenated to obtain a new braid and this simple rule defines the group composition. Pictorially, a braid can be represented as $N$ strands entering a box and being acted on by a set of braid moves that are defined up to a topological equivalence or, equivalently, to the so called Reidemeister moves \cite{KauffmanBook1}. The $N$ strands exit the box at the same locations they entered. The composition of two braids reduces to stacking the boxes, erasing the middle segment and compressing the resulting rectangle until becomes the box. The unit of the group consists of $N$ parallel strands. All these are illustrated in Fig.~\ref{Fig:PictorialBraidGroup}. 

\begin{figure}[t]
\center
\includegraphics[width=0.7\linewidth]{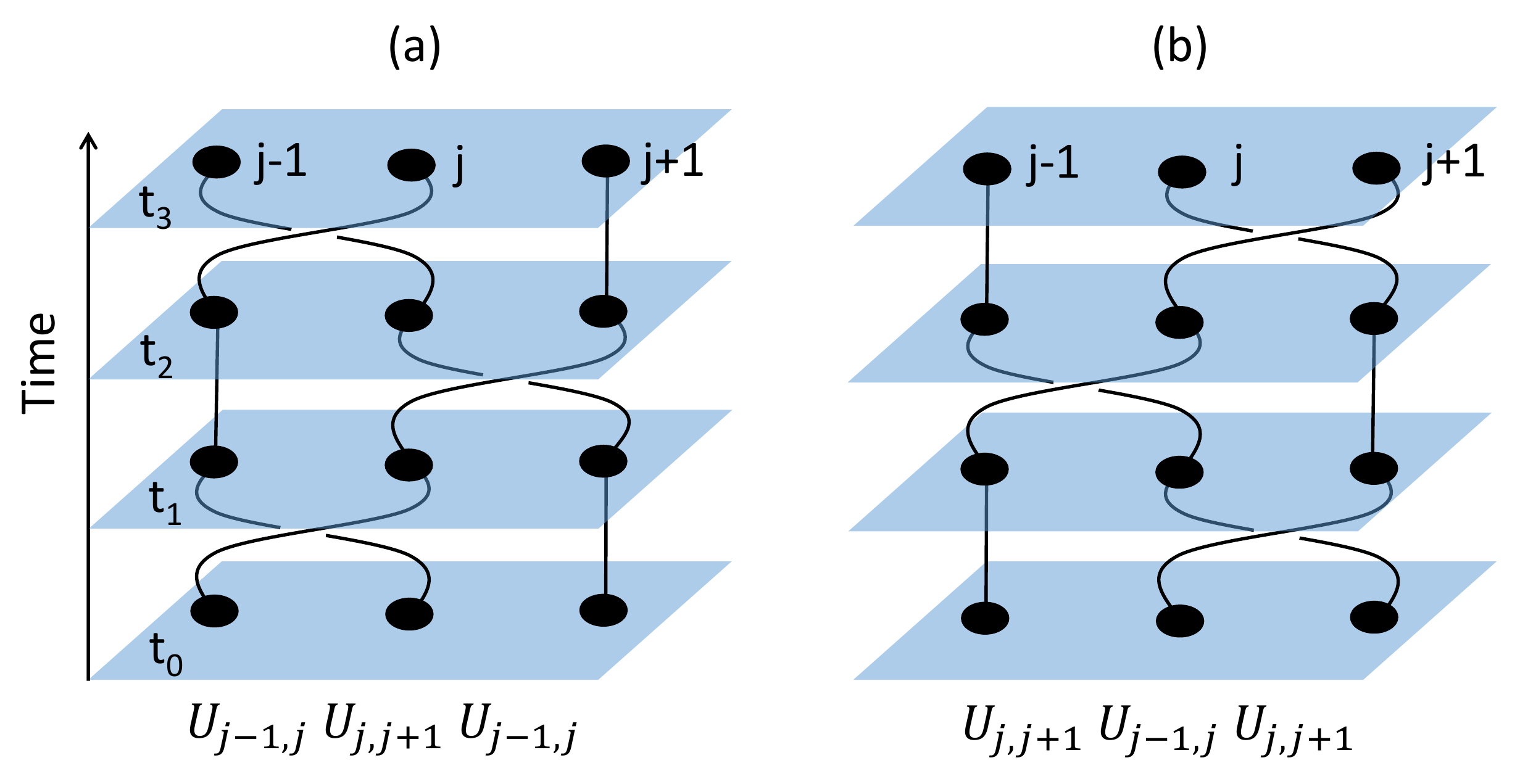}
\caption{\small Sequences of adiabatic cycles which exchange flux tubes. If the resulting unitary transformations depend solely on the topology of the world lines, then $U_{j-1,j}U_{j,j+1}U_{j-1,j}$ coincides with $U_{j,j+1}U_{j-1,j}U_{j,j+1}$ because the (a) and (b) world lines are topologically equivalent. Since these are the defining relations of the braid group \cite{BraidBook}, the adiabatic exchanges supply a unitary representation of the abstract braid group. }
\label{Fig:AdiabaticBraidGroup}
\end{figure}

\vspace{0.2cm}

While the braid group $\Bb_N$ has an infinite number of elements, it has a finite number of generators. These are $\beta_{j,j+1}$, $j=1,\ldots,N-1$, where $\beta_{j,j+1}$ weaves the strands $j$ and $j+1$. An illustration of such generator and its inverse can be found in Fig.~\ref{Fig:PictorialBraidGroup}. In terms of these generators, the braid group can be defined as the group generated by the $\beta_{j,j+1}$'s together with the relations:
\begin{align}\label{Eq:BRelations}
& \beta_{j,j+1} \beta_{j+1,j+2} \beta_{j,j+1} = \beta_{j+1,j+2} \beta_{j.j+1} \beta_{j+1,j+2}, \quad j=1,\ldots,N-2, \\
& \beta_{j,j+1} \beta_{k,k+1} = \beta_{k,k+1} \beta_{j,j+1}, \quad j,k=1,\ldots,N-1, \quad |j-k| \geq 2.
\end{align}
This is known as the Artin presentation of the braid group \cite{ArtinAM1947}. An element of the group is the equivalence class w.r.t. \eqref{Eq:BRelations} of a word like:
\begin{equation}
\beta_{i_1,i_1+1}^{n_1} \beta_{i_2,i_2+1}^{n_2} \ldots \beta_{i_k,i_k+1}^{n_k},
\end{equation}
where the indices $i_\alpha$ are drawn from $\{1,\ldots,N-1\}$ and the powers $n_\alpha$ can be any integer number. Note that the size of a word can grow indefinitely.

\begin{figure}[b]
\centering
\includegraphics[width=0.8\linewidth]{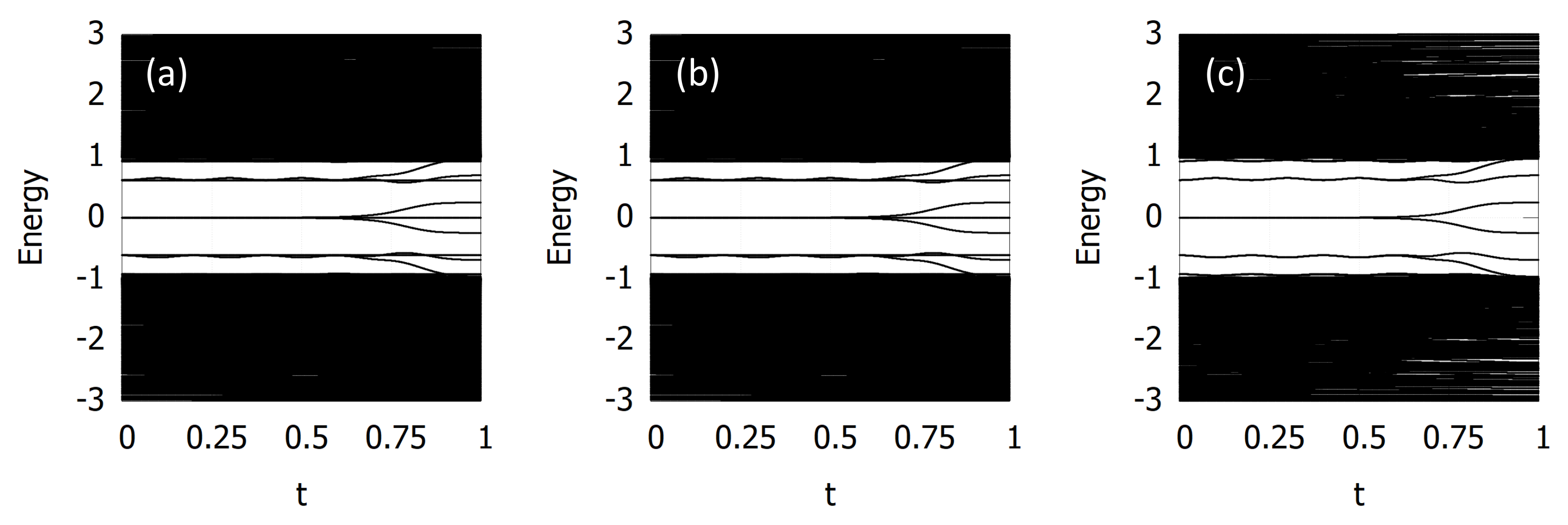}
\caption{\small Evolution of the spectrum during (a) $(-\pi, -\pi)$, (b) $(+\pi, +\pi)$ and (c) $(+\pi, -\pi)$ fusions. The tubes are separated at $t=0$ and overlap at $t=1$ and for $(-\pi, -\pi)$ and $(+\pi, +\pi)$ fusion, two pairs of $(+\pi, -\pi)$ tubes were created. The zero modes seen after fusion correspond to the un-fused flux-tubes.}
\label{Fig:FusionSpectrum}
\end{figure}

\subsection{Adiabatic representation of the braid group} 

\vspace{0.2cm}

We consider now the physical processes illustrated in Fig.~\ref{Fig:AdiabaticBraidGroup}. They consists of adiabatic displacements of (+)-flux-tubes which result in permutations of the tubes. Note that at each marked times, $t_0$, \ldots, $t_3$, the system is in the same exact configuration. Let $H_t$ and $U(t,t')$ denote the time-dependent Hamiltonian and the unitary time evolution induced by the processes in Fig.~\ref{Fig:AdiabaticBraidGroup}(a), and $H'_t$ and $U'(t,t')$ be the same objects for the processes in Fig.~\ref{Fig:AdiabaticBraidGroup}(b). Clearly:
\begin{equation}
U(t_1,t_0) = U(t_3,t_2) = U'(t_2,t_1)
\end{equation}
and these identical unitary operators will be denoted as $U_{j-1,j}$. Similarly:
\begin{equation}
U(t_2,t_1) = U'(t_1,t_0) = U'(t_3,t_2),
\end{equation}
and these identical unitary operators will be denoted as $U_{j,j+1}$. 
Since after each of these time-intervals the system returns in the original configuration $x_0$, both $U_{j-1,j}$ and $U_{j,j+1}$ act on $P_{x_0}\Hh$. Furthermore, it follows straight from \eqref{Eq:GroupProperty} that:
\begin{equation}
U(t_3,t_0)=U(t_3,t_2)U(t_2,t_1)U(t_1,t_0) = U_{j-1,j}U_{j,j+1} U_{j-1,j},
\end{equation} 
and similarly:
\begin{equation}
U'(t_3,t_0)=U'(t_3,t_2)U'(t_2,t_1)U'(t_1,t_0) = U_{j,j+1}U_{j-1,j} U_{j,j+1}.
\end{equation} 

\vspace{0.2cm}

In general, there is no reason to believe that $U(t_3,t_0)=U'(t_3,t_0)$ because, after all, the physical processes in panels (a) and (b) of Fig.~\ref{Fig:AdiabaticBraidGroup} are distinct. However, recall that the adiabatic evolutions are geometric in nature and, since the zero-modes are concentrated near the flux-tubes,  the world lines can be deformed without any physical consequences as long as the flux-tubes are kept far apart from each other. If that is the case, then the world lines in panels (a) and (b) of Fig.~\ref{Fig:AdiabaticBraidGroup} can be deformed into each other without affecting the final outcomes. The conclusion is that:
\begin{equation}\label{Eq:Relation}
U_{j-1,j}U_{j,j+1} U_{j-1,j} = U_{j,j+1}U_{j-1,j} U_{j,j+1},
\end{equation}
and these are exactly the defining relations \eqref{Eq:BRelations} of the braid group. The conclusion is that the adiabatic evolutions supply a representation of the braid group on $P_{x_0}\Hh$, which we call the adiabatic representation of the braid group.

\vspace{0.2cm}

The reason why one can expect a non-trivial representation of the braid group is because the parameter space of the flux-tubes is punctured. Indeed, in Fig.~\ref{Fig:FusionSpectrum} we illustrate the behavior of the spectrum as pairs of flux tubes are fused. As one can see, the degeneracy is lifted and, as such, the adiabatic theorem, in the form we presented, does not apply anymore because dynamical terms must be also taken into account. As shown in \cite{ProdanPRB2009}, a Berry curvature can be associated to the adiabatic deformations and this curvature is usually concentrated near such singularities, leading in general to non-trivial monodromies.

\section{Braiding the Zero Modes}
\label{Sec:Braiding0s}

\begin{figure}[t]
\centering
\includegraphics[width=0.8\linewidth]{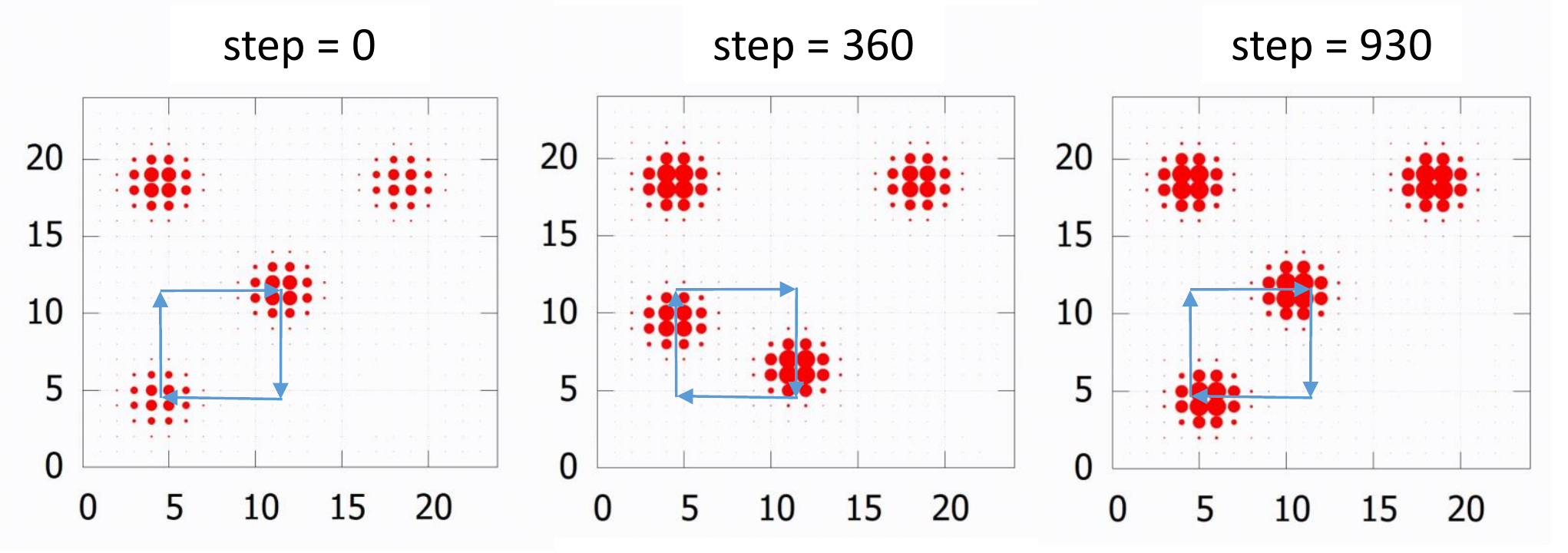}
\caption{\small Plots of $\sum_{\alpha=1,2} \langle \bm n,\alpha|P_{x_t}|\bm n,\alpha \rangle$ defined in \eqref{Eq:ZProj} as function of lattice position $\bm n$, at three instances along the adiabatic cycle shown by the arrows. The cycle was discretized in 1000 steps. The value at a point $\bm n$ of the $25 \times 25$ lattice is proportional with the size of the disk centered at that position.}
\label{Fig:2U12Braiding}
\end{figure} 

In this section we compute explicitly the braid matrices and confirm that they supply an adiabatic representations of the braid group. In all instances, the flux-tubes are nucleated in pairs of $\Phi=\pm \pi$ but the (-) flux-tubes will be fixed and only the (+) flux-tubes will be braided. 

 \begin{figure}[t]
\centering
\includegraphics[width=\linewidth]{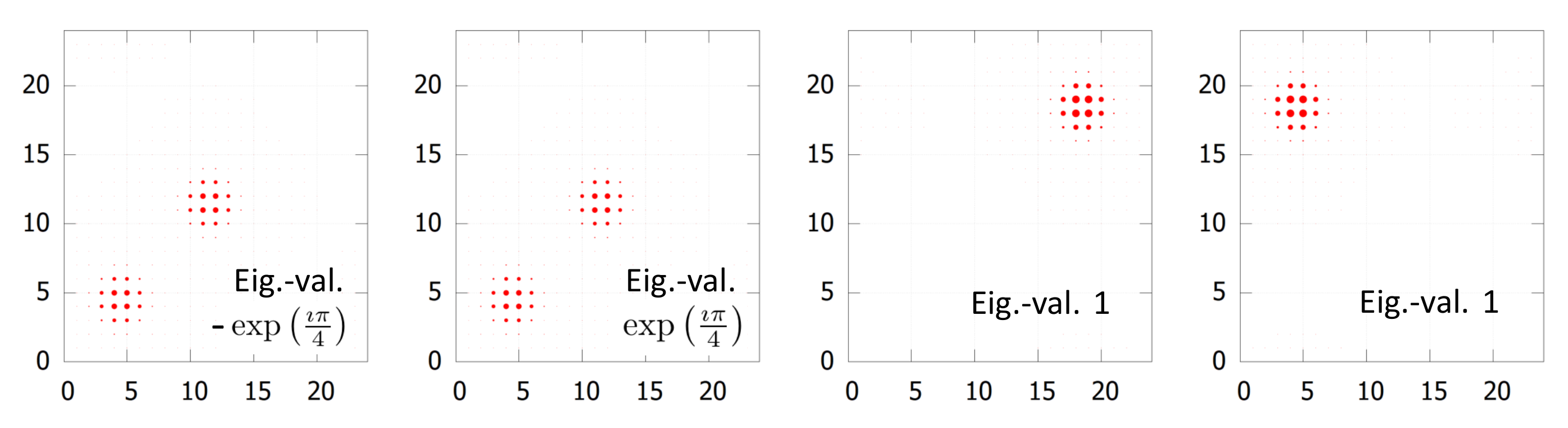}
\caption{\small Plots of the eigen-vectors of $U_{12}$ with the latter computed with the process shown in Fig.~\ref{Fig:2U12Braiding}. The panels show the amplitude $|\psi(\bm n)|^2$ of the vectors as function of position $\bm n$ on the lattice. The corresponding eigen-values are specified in each panel.}
\label{Fig:2U12Vectors}
\end{figure} 

\subsection{Two (+) flux-tubes ($\Bb_2$)}
\label{Sec:AdiabaticB2} 

\vspace{0.2cm}

The exchange route of the (+) flux-tubes as well as the positions of the fixed (-) flux tubes are shown in Fig.~\ref{Fig:2U12Braiding}. The dimension of the zero modes space is 4. The spatial profiles of the zero-modes $\varphi_j$ trapped by the four flux-tubes are conveniently captured by the diagonal part of $P_x$, the projection onto the zero modes space:
\begin{equation}\label{Eq:ZProj}
\sum_{\alpha=1,2} \langle \bm n,\alpha|P_x|\bm n,\alpha \rangle = \sum_{j=1}^4 |\varphi_j(\bm n,\alpha)|^2,
\end{equation}
which is what is plotted in Fig.~\ref{Fig:2U12Braiding}. As one can see, the zero-modes remain relatively well separated during the exchange operation. Furthermore, since the flux-tubes that are exchanged carry the same flux, the Hamiltonian returns to its initial value, hence the adiabatic cycle is closed.

\vspace{0.2cm}

The monodromy $U_{12}$ was computed with Eq.~\ref{Eq:PracticalUa} and, for a discretization of $K=5 \times 10^4$, its spectrum was found to be:
\begin{equation}\label{Eq:2U12Spec}
{\rm Spec}(U_{12}) = \big \{-\exp\big(\tfrac{\imath \pi}{4}\big),\exp\big ( \tfrac{\imath \pi}{4}\big ),1,1 \big \} \pm 1\%. 
\end{equation} 
The corresponding eigen-vectors are reported in Fig.~\ref{Fig:2U12Vectors}. As expected, the trivial eigenvalues 1 correspond to the zero modes trapped by the (-) flux-tubes, which do not participate in the braiding. The eigen-vectors corresponding to the eige-nvalues $\pm\exp\big(\tfrac{\imath \pi}{4}\big)$ are shared between the (+) flux tubes and they are mapped into each other by the particle-hole symmetry operation \eqref{Eq:PHSymmetry}.

\vspace{0.2cm}

The above numerical results indicate that when a flux tube is circled around another, the resulting adiabatic monodromy is $\pm \imath I$. We have verified this statement, directly, for the process described in Fig.~\ref{Fig:UTest}. Hence, the flux-tube must circle four times for the monodromy to return to identity. The conclusion is that the adiabatic representation supply a cyclic representation of $\Bb_2$ of period 8: $U_{12}^8 = I$. 

 \begin{figure}[t]
\centering
\includegraphics[width=0.6\linewidth]{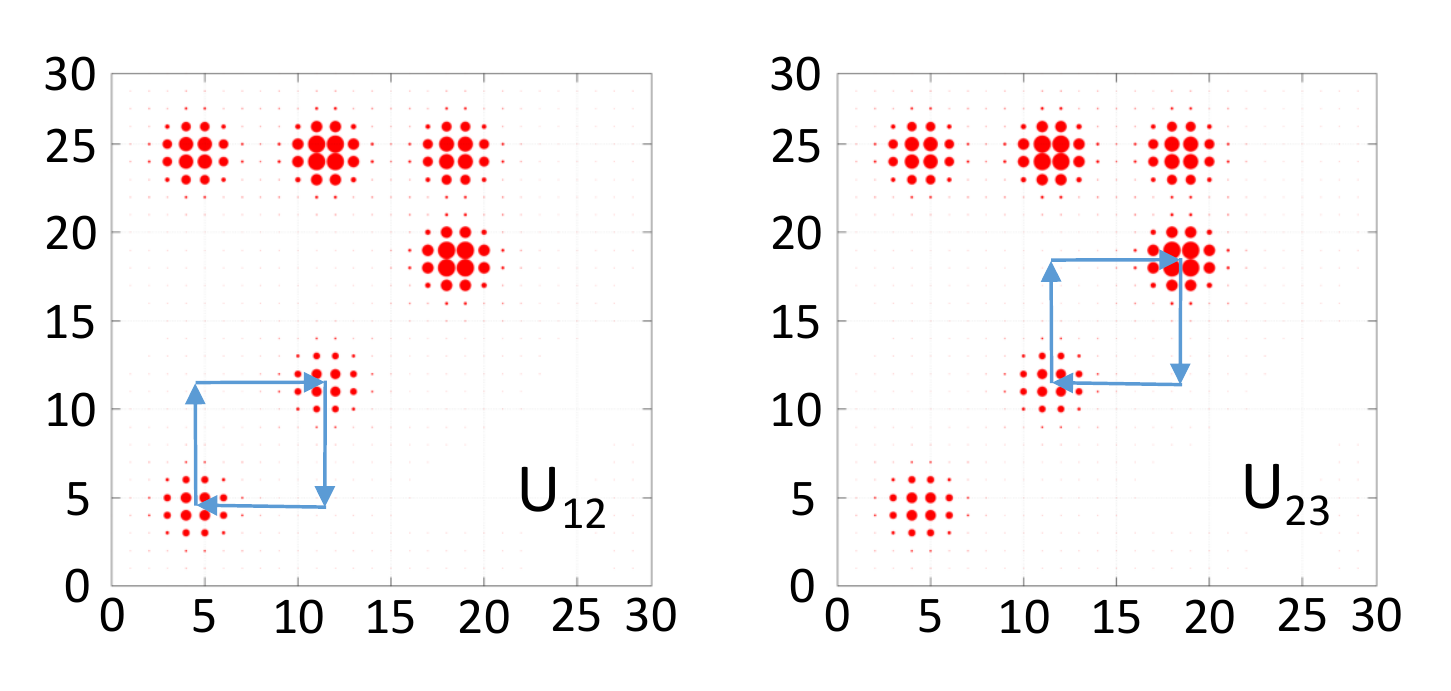}
\caption{\small The initial configuration of the flux-tubes together with the world-lines for the exchange matrices marked in each panel. The initial configuration is represented by the plot of $\sum_{\alpha=1,2} \langle \bm n,\alpha|P_{x_0} | \bm n,\alpha\rangle$ as function of $\bm n$ on a $30\times 30$ lattice.}
\label{Fig:3UBraidings}
\end{figure}

\subsection{Three (+) flux-tubes ($\Bb_3$)} 
\label{Sec:AdiabaticB3}

\vspace{0.2cm}

\begin{figure}[t]
  	\centering
 	\includegraphics[width=0.8\linewidth]{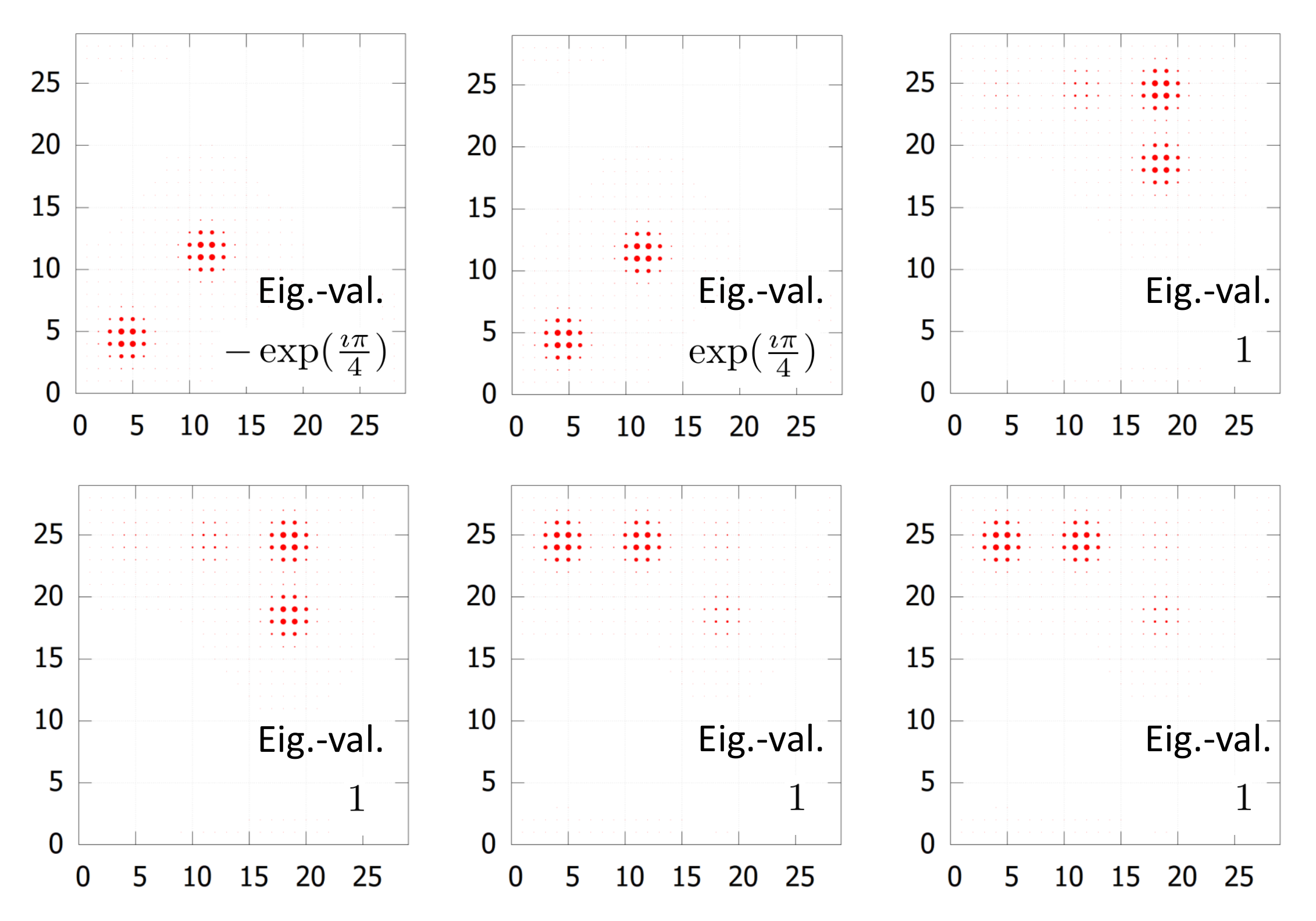}
 	\caption{\small The eigen-vectors of $U_{12}$ for the eigen-values specified for each panel. }
 	\label{Fig:3U12Vectors}
 \end{figure}

\begin{figure}[t]
 	\centering
 	\includegraphics[width=0.8\linewidth]{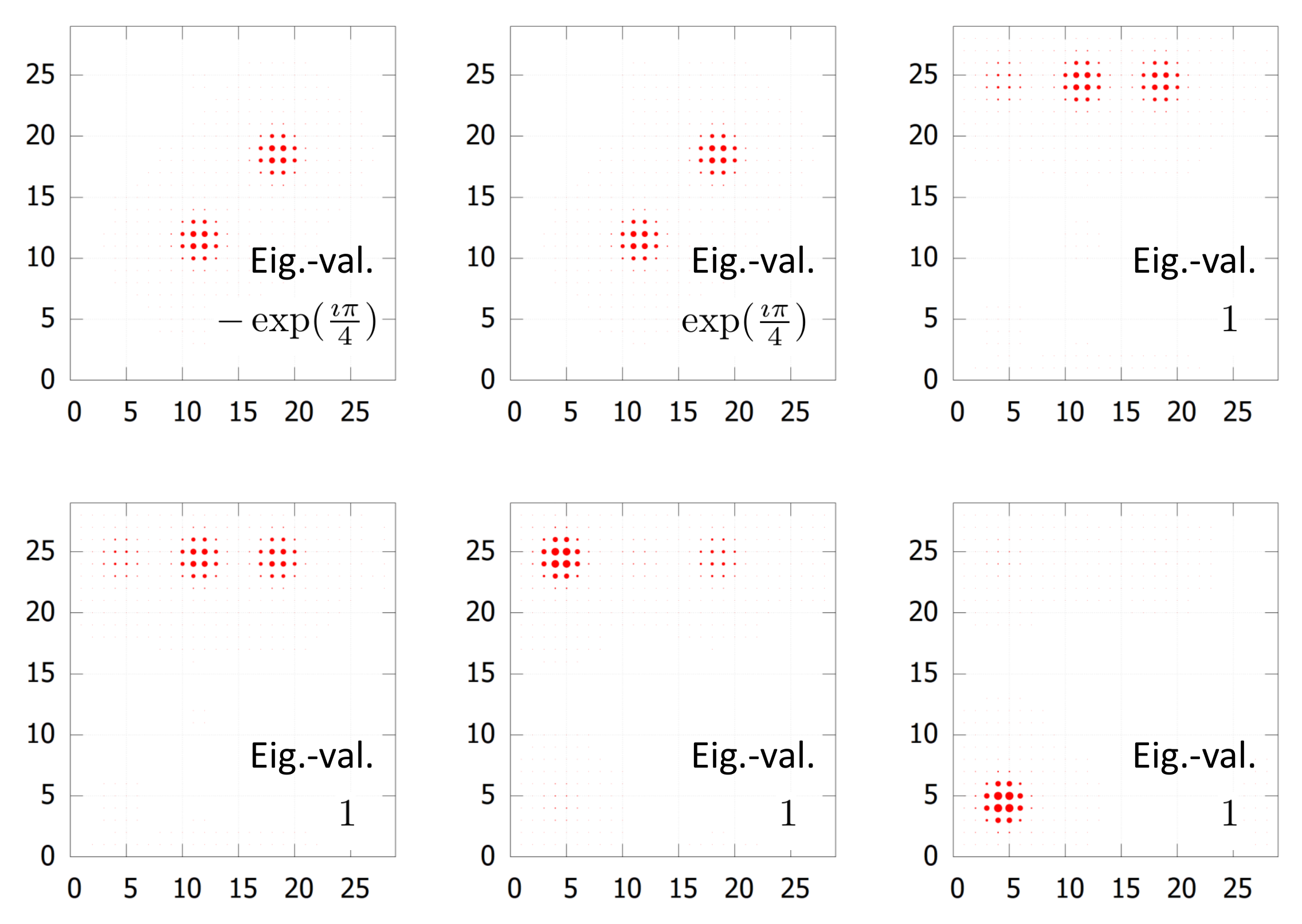}
 	\caption{\small  Eigen-vectors of $U_{23}$ for the eigen-values specified for each panel.}
 	\label{Fig:3U23Vectors}
 \end{figure}

The initial configurations and the exchange routes of flux-tubes are shown in Fig.~\ref{Fig:3UBraidings} and the monodromies $U_{12}$ and $U_{23}$ were again computed with Eq.~\ref{Eq:PracticalUa}. The eigenvalues of both monodromies were numerically found to be:
\begin{equation}\label{Eq:SpecU12U23}
{\rm Spec}(U_{12}) = {\rm Spec}(U_{23}) = \big \{-\exp(\tfrac{\imath\pi}{4}),\exp(\tfrac{\imath\pi}{4}),1,1,1,1 \big \} \pm 1\%, 
\end{equation}
when the adiabatic cycle was discretized in a $K=5 \times 10^4$ number of steps. The corresponding eigen-vectors are shown in Fig.~\ref{Fig:3U12Vectors} for $U_{12}$ and in Fig.~\ref{Fig:3U23Vectors} for $U_{23}$. At this point we computed the physical representation of the generators of the braid group and we can verify directly the fundamental relation \eqref{Eq:Relation}. Numerically, when the cycles were discretized in $K=5 \times 10^4$ steps, we obtained:
\begin{equation}
U_{12}U_{23}U_{12} - U_{23}U_{12}U_{23} = 0\pm 0.01,
\end{equation}
which confirms that the adiabatic braids supply a representation of the braid group. The spectra of these unitary transformations was found to be:
\begin{equation}\label{Eq:SpecU123}
{\rm Spec}(U_{12}U_{23}U_{12}) =\{-\imath,\ \imath,\ \imath,\ 1,\ 1,\ 1\} \pm 1\%,
\end{equation}
and similarly for ${\rm Spec}(U_{23}U_{12}U_{23})$.

 \vspace{0.2cm}
 
 Since three of the flux-tubes are mere spectators to the braidings, we expect that three of the eigen-vectors corresponding to trivial eigenvalues 1 of $U_{12}$ and $U_{23}$ to be pinned at these (-) flux-tubes and to be irrelevant for the braiding process. This is not clear from Figs.~\ref{Fig:3U12Vectors} and \ref{Fig:3U23Vectors} and this is because the computer out-putted arbitrary linear combinations of the four eige-vectors corresponding to the degenerate eigenvalue 1. However, to confirm our supposition, we have verified that the two eigen-vectors of $U_{12}$ and the two eige-vectors of $U_{23}$, corresponding to the non-trivial eigen-values, span a 3-dimensional linear space, {\it i.e.} that there is one and only one linear dependency between them. Furthermore, this 3-dimensional linear space coincides with the 3-dimensional linear space spanned by the three eigen-vectors of $U_{23}U_{12}U_{23}$ corresponding to the three non-trivial eigenvalues. The important conclusion is that the adiabatic representation of the braid group revealed by our calculations is an irreducible representation of dimension 3.
 
\section{The Resulting Irreducible Representation of $\Bb_N$}   

The numerical data supplied in the previous section enable us to identify the irreducible representation of the braid group for a generic number of flux tubes. This representation is described first in a matrix form and then in a diagramatic form. 

\subsection{The matrix presentation}
\label{Sec:MatrixPres}

\vspace{0.2cm}

Consider the following matrices $N\times N$ complex matrices:
\begin{equation}\label{Eq:Qs}
Q_{j,j+1} =
\begin{pmatrix}
\ddots & \vdots & \vdots & \vdots & \vdots   & \iddots \\
\ldots & \gamma & 0        &  0         &  0           & \ldots \\
\ldots &  0          &  0        & \alpha &  0           & \ldots \\
\ldots &  0          &  \beta & 0         &  0           & \ldots \\
\ldots &  0          &  0        & 0         &  \gamma & \ldots \\
\iddots  & \vdots & \vdots  & \vdots  & \vdots  & \ddots
\end{pmatrix},
\end{equation}
where the $2 \times 2$ square matrix containing the coefficients $\alpha$ and $\beta$ sits at rows $j$ and $j+1$, as well as at columns $j$ and $j+1$. The $\gamma$'s fill in the diagonal all the way to the upper-left and lower-right corners. We can let the index $j$ run from $1$ to $N-1$, hence there are $N-1$ such $Q$ matrices. The adjoint matrices are:
\begin{equation}
Q_{j,j+1}^\ast =
\begin{pmatrix}
\ddots & \vdots & \vdots & \vdots & \vdots   & \iddots \\
\ldots & \gamma^\ast & 0        &  0         &  0           & \ldots \\
\ldots &  0          &  0        & \beta^\ast &  0           & \ldots \\
\ldots &  0          &  \alpha^\ast & 0         &  0           & \ldots \\
\ldots &  0          &  0        & 0         &  \gamma^\ast & \ldots \\
\iddots  & \vdots & \vdots  & \vdots  & \vdots  & \ddots
\end{pmatrix},
\end{equation}
 and a direct computation gives:
\begin{equation}
Q_{j,j+1} Q_{j,j+1}^\ast =
\begin{pmatrix}
\ddots & \vdots & \vdots & \vdots & \vdots   & \iddots \\
\ldots & |\gamma|^2 & 0        &  0         &  0           & \ldots \\
\ldots &  0          &  |\alpha|^2        & 0  &  0           & \ldots \\
\ldots &  0          &  0 & |\beta|^2         &  0           & \ldots \\
\ldots &  0          &  0        & 0         & |\gamma|^2 & \ldots \\
\iddots  & \vdots & \vdots  & \vdots  & \vdots  & \ddots
\end{pmatrix},
\end{equation}
as well as:
\begin{equation}
Q_{j,j+1}^\ast Q_{j,j+1} =
\begin{pmatrix}
\ddots & \vdots & \vdots & \vdots & \vdots   & \iddots \\
\ldots & |\gamma|^2 & 0        &  0         &  0           & \ldots \\
\ldots &  0          &  |\beta|^2        & 0  &  0           & \ldots \\
\ldots &  0          &  0 & |\alpha|^2         &  0           & \ldots \\
\ldots &  0          &  0        & 0         & |\gamma|^2 & \ldots \\
\iddots  & \vdots & \vdots  & \vdots  & \vdots  & \ddots
\end{pmatrix}.
\end{equation}
As such, the matrices are unitary if we force the complex coefficients $\alpha$, $\beta$ and $\gamma$ to lie on the unit circle. Furthermore:
\begin{equation}
Q_{j,j+1} Q_{j+1,j+2} Q_{j,j+1} =
\begin{pmatrix}
\ddots & \vdots      & \vdots                   & \vdots                        & \vdots                    & \iddots \\
\ldots & \gamma^3 & 0                           &  0                                &  0                            & \ldots \\
\ldots &  0              &  0                           & 0                                 &  \gamma \alpha^2 & \ldots \\
\ldots &  0              &  0                           & \gamma \alpha \beta &  0                           & \ldots \\
\ldots &  0              &  \gamma \beta^2  & 0                                  & 0                           & \ldots \\
\ldots &  0              &  0                          & 0                                   & \gamma^3            & \ldots \\
\iddots  & \vdots   & \vdots                   & \vdots                          & \vdots                  & \ddots
\end{pmatrix},
\end{equation}
where the $3\times 3$ off-diagonal matrix sits at rows $j$, $j+1$, $j+2$ and columns $j$, $j+1$, $j+2$, and a direct computation will show that:
\begin{equation}
Q_{j,j+1} Q_{j+1,j+2} Q_{j,j+1} = Q_{j+1,j+2} Q_{j,j+1} Q_{j+1,j+2},
\end{equation}
as well as:
\begin{equation}
Q_{j,j+1} Q_{j',j'+1} = Q_{j',j'+1} Q_{j,j+1},
\end{equation}
if $|j-j'|\geq 2$. Comparing with \eqref{Eq:BRelations}, we conclude that the $Q$-matrices supply an $N \times N$ representation of the braid group $\Bb_N$. 

\subsection{Connection with the adiabatic physical representation}\label{Sec:PhysicalRep}

\vspace{0.2cm}

Specializing for $N=3$, we have:
\begin{equation}\label{Eq:QN3}
Q_{12} =
\begin{pmatrix}
0        & \alpha &  0  \\
\beta & 0         &  0  \\
0        & 0         &  \gamma  \\
\end{pmatrix}, \quad 
Q_{23} =
\begin{pmatrix}
\gamma & 0        &  0          \\
0          &  0        & \alpha    \\
0          &  \beta & 0           \\
\end{pmatrix}.
\end{equation}
Guided by the numerical results from section~\ref{Sec:Braiding0s}, we make the choice:
\begin{equation}\label{Eq:AlphaBeta}
\alpha = e^{-\frac{\imath \pi}{4}}, \quad \beta = - e^{-\frac{\imath \pi}{4}}, \quad \gamma=1.
\end{equation}
and find:
\begin{equation}
{\rm Spec}(Q_{12}) = {\rm Spec}(Q_{12}) = \big \{ - e^{\frac{\imath \pi}{4}}, e^{\frac{\imath \pi}{4}}, 1 \big \}
\end{equation}
as well as:
\begin{equation}
{\rm Spec}(Q_{12}Q_{23}Q_{12}) = {\rm Spec}(Q_{23}Q_{12}Q_{23})=\big \{ - \imath, \imath, \imath \big \},
\end{equation}
in full agreement with \eqref{Eq:SpecU12U23} and \eqref{Eq:SpecU123}.

\subsection{A diagramatic presentation}

\vspace{0.2cm}

A single-strand $N$-diagram is a box with $N$-marked points at the bottom edge and $N$-marked points on the top edge, and a smooth line connecting one marked point at the base with one marked point at the top. The smooth line exists and dives normally to the edges. A few examples for $N=3$ are supplied below:
 \begin{equation}\label{Eq:B3Diags}
 \begin{matrix}
 \begin{tikzpicture}
 \draw[color=red] (0,0) rectangle (1.2,0.7);
 \filldraw (0.3,0) circle [radius=1pt];
 \filldraw (0.6,0) circle [radius=1pt];
 \filldraw (0.9,0) circle [radius=1pt];
 \filldraw (0.3,0.7) circle [radius=1pt];
 \filldraw (0.6,0.7) circle [radius=1pt];
 \filldraw (0.9,0.7) circle [radius=1pt];
 \draw (0.3,0) .. controls +(0,0.2) and +(0,-0.2) .. (0.9,0.7);
 \end{tikzpicture}
 \end{matrix}
 \begin{matrix}
 , \ \ 
 \end{matrix}
  \begin{matrix}
 \begin{tikzpicture}
 \draw[color=red] (0,0) rectangle (1.2,0.7);
 \filldraw (0.3,0) circle [radius=1pt];
 \filldraw (0.6,0) circle [radius=1pt];
 \filldraw (0.9,0) circle [radius=1pt];
 \filldraw (0.3,0.7) circle [radius=1pt];
 \filldraw (0.6,0.7) circle [radius=1pt];
 \filldraw (0.9,0.7) circle [radius=1pt];
 \draw (0.3,0) -- (0.3,0.7);
 \end{tikzpicture}
 \end{matrix}
 \begin{matrix}
 , \ \ 
 \end{matrix}
  \begin{matrix}
 \begin{tikzpicture}
 \draw[color=red] (0,0) rectangle (1.2,0.7);
 \filldraw (0.3,0) circle [radius=1pt];
 \filldraw (0.6,0) circle [radius=1pt];
 \filldraw (0.9,0) circle [radius=1pt];
 \filldraw (0.3,0.7) circle [radius=1pt];
 \filldraw (0.6,0.7) circle [radius=1pt];
 \filldraw (0.9,0.7) circle [radius=1pt];
 \draw (0.6,0) .. controls +(0,0.2) and +(0,-0.2) .. (0.3,0.7);
 \end{tikzpicture}
 \end{matrix}
 \end{equation}
For a generic $N$, there are $N^2$ such single-strand diagrams.

\vspace{0.2cm} 
 
Two single-strand diagrams can be composed in the following way. One stacks the diagrams on top of each other, respecting the order, and, if the reunion of the paths results in a smooth curve then that is the result of the composition. If not, then the composition is zero. Below are some example:
 \begin{equation}\label{Eq:CExample1}
   \begin{matrix}
 \begin{tikzpicture}
 \draw[color=red] (0,0) rectangle (1.2,0.7);
 \filldraw (0.3,0) circle [radius=1pt];
 \filldraw (0.6,0) circle [radius=1pt];
 \filldraw (0.9,0) circle [radius=1pt];
 \filldraw (0.3,0.7) circle [radius=1pt];
 \filldraw (0.6,0.7) circle [radius=1pt];
 \filldraw (0.9,0.7) circle [radius=1pt];
 \draw (0.9,0) .. controls +(0,0.2) and +(0,-0.2) .. (0.6,0.7);
 \end{tikzpicture}
 \end{matrix}
 \begin{matrix}
 \ \cdot \
 \end{matrix}
 \begin{matrix}
 \begin{tikzpicture}
 \draw[color=red] (0,0) rectangle (1.2,0.7);
 \filldraw (0.3,0) circle [radius=1pt];
 \filldraw (0.6,0) circle [radius=1pt];
 \filldraw (0.9,0) circle [radius=1pt];
 \filldraw (0.3,0.7) circle [radius=1pt];
 \filldraw (0.6,0.7) circle [radius=1pt];
 \filldraw (0.9,0.7) circle [radius=1pt];
 \draw (0.3,0) .. controls +(0,0.2) and +(0,-0.2) .. (0.9,0.7);
 \end{tikzpicture}
 \end{matrix}
 \begin{matrix}
\ = \ 
 \end{matrix}
 \begin{matrix}
 \begin{tikzpicture}
 \draw[color=red] (0,0) rectangle (1.2,0.7);
 \filldraw (0.3,0) circle [radius=1pt];
 \filldraw (0.6,0) circle [radius=1pt];
 \filldraw (0.9,0) circle [radius=1pt];
 \filldraw (0.3,0.7) circle [radius=1pt];
 \filldraw (0.6,0.7) circle [radius=1pt];
 \filldraw (0.9,0.7) circle [radius=1pt];
 \draw (0.3,0) .. controls +(0,0.2) and +(0,-0.2) .. (0.9,0.7);
 \draw[color=red] (0,0.7) rectangle (1.2,1.4);
 \filldraw (0.3,0.7) circle [radius=1pt];
 \filldraw (0.6,0.7) circle [radius=1pt];
 \filldraw (0.9,0.7) circle [radius=1pt];
 \filldraw (0.3,1.4) circle [radius=1pt];
 \filldraw (0.6,1.4) circle [radius=1pt];
 \filldraw (0.9,1.4) circle [radius=1pt];
 \draw (0.9,0.7) .. controls +(0,0.2) and +(0,-0.2) .. (0.6,1.4);
 \end{tikzpicture}
 \end{matrix}
  \begin{matrix}
 \ = \ 
 \end{matrix}
 \begin{matrix}
 \begin{tikzpicture}
 \draw[color=red] (0,0) rectangle (1.2,0.7);
 \filldraw (0.3,0) circle [radius=1pt];
 \filldraw (0.6,0) circle [radius=1pt];
 \filldraw (0.9,0) circle [radius=1pt];
 \filldraw (0.3,0.7) circle [radius=1pt];
 \filldraw (0.6,0.7) circle [radius=1pt];
 \filldraw (0.9,0.7) circle [radius=1pt];
 \draw (0.3,0) .. controls +(0,0.2) and +(0,-0.2) .. (0.6,0.7);
  \end{tikzpicture}
 \end{matrix}
 \begin{matrix}
 \, ,
 \end{matrix}
 \end{equation}
 and:
  \begin{equation}\label{Eq:CExample1}
   \begin{matrix}
 \begin{tikzpicture}
 \draw[color=red] (0,0) rectangle (1.2,0.7);
 \filldraw (0.3,0) circle [radius=1pt];
 \filldraw (0.6,0) circle [radius=1pt];
 \filldraw (0.9,0) circle [radius=1pt];
 \filldraw (0.3,0.7) circle [radius=1pt];
 \filldraw (0.6,0.7) circle [radius=1pt];
 \filldraw (0.9,0.7) circle [radius=1pt];
 \draw (0.6,0) .. controls +(0,0.2) and +(0,-0.2) .. (0.3,0.7);
 \end{tikzpicture}
 \end{matrix}
 \begin{matrix}
 \ \cdot \
 \end{matrix}
 \begin{matrix}
 \begin{tikzpicture}
 \draw[color=red] (0,0) rectangle (1.2,0.7);
 \filldraw (0.3,0) circle [radius=1pt];
 \filldraw (0.6,0) circle [radius=1pt];
 \filldraw (0.9,0) circle [radius=1pt];
 \filldraw (0.3,0.7) circle [radius=1pt];
 \filldraw (0.6,0.7) circle [radius=1pt];
 \filldraw (0.9,0.7) circle [radius=1pt];
 \draw (0.9,0) .. controls +(0,0.2) and +(0,-0.2) .. (0.3,0.7);
 \end{tikzpicture}
 \end{matrix}
 \begin{matrix}
\ = \ 
 \end{matrix}
 \begin{matrix}
 \begin{tikzpicture}
 \draw[color=red] (0,0) rectangle (1.2,0.7);
 \filldraw (0.3,0) circle [radius=1pt];
 \filldraw (0.6,0) circle [radius=1pt];
 \filldraw (0.9,0) circle [radius=1pt];
 \filldraw (0.3,0.7) circle [radius=1pt];
 \filldraw (0.6,0.7) circle [radius=1pt];
 \filldraw (0.9,0.7) circle [radius=1pt];
 \draw (0.9,0) .. controls +(0,0.2) and +(0,-0.2) .. (0.3,0.7);
 \draw[color=red] (0,0.7) rectangle (1.2,1.4);
 \filldraw (0.3,0.7) circle [radius=1pt];
 \filldraw (0.6,0.7) circle [radius=1pt];
 \filldraw (0.9,0.7) circle [radius=1pt];
 \filldraw (0.3,1.4) circle [radius=1pt];
 \filldraw (0.6,1.4) circle [radius=1pt];
 \filldraw (0.9,1.4) circle [radius=1pt];
 \draw (0.6,0.7) .. controls +(0,0.2) and +(0,-0.2) .. (0.3,1.4);
 \end{tikzpicture}
 \end{matrix}
  \begin{matrix}
 \ = 0 .
 \end{matrix}
 \end{equation}
As opposed to the link diagrams, not every single-strand diagram has an inverse. In fact there is no single-strand diagram that can play the role of unity, hence the set of these diagrams together with the composition form only a semi-group. 

\vspace{0.2cm}

This semi-group is not very useful by himself and we expand to an algebra $T^{(1)}_N$, whose elements are formal series:
\begin{equation}\label{Eq:TLElement}
a = \sum_{j} a_j \, \boxed{D_j} \, ,
\end{equation}
where $\boxed{D_j}$'s are single strand $N$-diagrams and the coefficients $a_j$'s are simple complex numbers. The addition and multiplication of the algebra are:
\begin{equation}
\Big ( \sum_{j} a_j \, \boxed{D_j} \Big ) + \Big (\sum_{j} a'_j \, \boxed{D_j}  \Big ) = \sum_{j} (a_j+a'_j) \, \boxed{D_j}
\end{equation}
and:
\begin{equation}
\Big ( \sum_{j} a_j \, \boxed{D_j} \Big ) \cdot \Big (\sum_{i} a'_i \, \boxed{D_j}  \Big ) = \sum_{j,i} a_j \, a'_i \ \boxed{\begin{matrix}D_j \\ D_i \end{matrix}} \, ,
\end{equation}
respectively.

\vspace{0.2cm}

The algebra $T^{(1)}_N$ does have a unit. For $N=3$, for example, the unit is supplied by:
\begin{equation}
\begin{matrix}
1 = \ 
\end{matrix}
\begin{matrix}
 \begin{tikzpicture}
 \draw[color=red] (0,0) rectangle (1.2,0.7);
 \filldraw (0.3,0) circle [radius=1pt];
 \filldraw (0.6,0) circle [radius=1pt];
 \filldraw (0.9,0) circle [radius=1pt];
 \filldraw (0.3,0.7) circle [radius=1pt];
 \filldraw (0.6,0.7) circle [radius=1pt];
 \filldraw (0.9,0.7) circle [radius=1pt];
 \draw (0.3,0) .. controls +(0,0.2) and +(0,-0.2) .. (0.3,0.7);
 \end{tikzpicture}
 \end{matrix}
 \begin{matrix}
 \ + \
 \end{matrix}
 \begin{matrix}
 \begin{tikzpicture}
 \draw[color=red] (0,0) rectangle (1.2,0.7);
 \filldraw (0.3,0) circle [radius=1pt];
 \filldraw (0.6,0) circle [radius=1pt];
 \filldraw (0.9,0) circle [radius=1pt];
 \filldraw (0.3,0.7) circle [radius=1pt];
 \filldraw (0.6,0.7) circle [radius=1pt];
 \filldraw (0.9,0.7) circle [radius=1pt];
 \draw (0.6,0) .. controls +(0,0.2) and +(0,-0.2) .. (0.6,0.7);
 \end{tikzpicture}
 \end{matrix}
 \begin{matrix}
 \ + \
 \end{matrix}
 \begin{matrix}
 \begin{tikzpicture}
 \draw[color=red] (0,0) rectangle (1.2,0.7);
 \filldraw (0.3,0) circle [radius=1pt];
 \filldraw (0.6,0) circle [radius=1pt];
 \filldraw (0.9,0) circle [radius=1pt];
 \filldraw (0.3,0.7) circle [radius=1pt];
 \filldraw (0.6,0.7) circle [radius=1pt];
 \filldraw (0.9,0.7) circle [radius=1pt];
 \draw (0.9,0) .. controls +(0,0.2) and +(0,-0.2) .. (0.9,0.7);
 \end{tikzpicture}
 \end{matrix}
\end{equation}
We can endow the algebra $T^{(1)}_N$ with a $\ast$-operation:
\begin{equation}
\Big ( \sum_{j} a_j \, \boxed{D_j} \Big )^\ast = \sum_{j} a_j^\ast \, \boxed{D_j}^{\, \ast},
\end{equation}
where $\boxed{D_j}^{\, \ast}$ for a single-strand diagram means reflection relative to the middle horizontal axis of the box. Furthermore, $T^{(1)}_N$ accepts a positive and faithful trace, {\it i.e.} an additive map $\Tt : T^{(1)}_N \rightarrow \CM$ such that $\Tt(gag^{-1}) = \Tt(a)$ for any $a$ from $T^{(1)}_N$ and $g$ from $GL(T_N)$ and $\Tt(a^\ast a) \geq 0$ for all $a \in T^{(1)}_N$, with equality only when $a=0$. Being a linear map, we only need to specify how $\Tt$ acts on the individual diagrams. The rule is that, when applied on a single-strand diagram, the trace gives $1$ if the strand is straight and $0$ otherwise. For example:
\begin{equation}
\begin{matrix}
\Tt
\end{matrix}
\begin{pmatrix} 
 \begin{tikzpicture}
 \draw[color=red] (0,0) rectangle (1.2,0.7);
 \filldraw (0.3,0) circle [radius=1pt];
 \filldraw (0.6,0) circle [radius=1pt];
 \filldraw (0.9,0) circle [radius=1pt];
 \filldraw (0.3,0.7) circle [radius=1pt];
 \filldraw (0.6,0.7) circle [radius=1pt];
 \filldraw (0.9,0.7) circle [radius=1pt];
 \draw (0.6,0) .. controls +(0,0.2) and +(0,-0.2) .. (0.6,0.7);
 \end{tikzpicture}
 \end{pmatrix}
 \begin{matrix}
  \ = 1, \quad  
 \end{matrix}
\begin{matrix}
\Tt
\end{matrix}
\begin{pmatrix} 
 \begin{tikzpicture}
 \draw[color=red] (0,0) rectangle (1.2,0.7);
 \filldraw (0.3,0) circle [radius=1pt];
 \filldraw (0.6,0) circle [radius=1pt];
 \filldraw (0.9,0) circle [radius=1pt];
 \filldraw (0.3,0.7) circle [radius=1pt];
 \filldraw (0.6,0.7) circle [radius=1pt];
 \filldraw (0.9,0.7) circle [radius=1pt];
 \draw (0.3,0) .. controls +(0,0.2) and +(0,-0.2) .. (0.9,0.7);
 \end{tikzpicture}
 \end{pmatrix}
 \begin{matrix}
  \ = 0.  
 \end{matrix}
 \end{equation}
Note that the trace is normalized such that $\Tt(1) = N$.

\vspace{0.2cm}

As for any algebra with a unit, there exists the group $GL(T_N)$ of all invertible elements from $T_N$. Below we supply an explicit group homomorphism 
\begin{equation}
\rho: \Bb_N \rightarrow GL(T_N),
\end{equation} 
via formal replacements of links by the linear combination:
\begin{equation}
\begin{matrix}
\begin{tikzpicture}
\draw[color=red] (0,0) -- (0.9,0);
\draw[color=red] (0,-0.7) -- (0.9,-0.7);
\braid[number of strands=2,width=0.3cm,height=0.7cm,border height=0cm] a_1;
 \filldraw (0.3,0) circle [radius=1pt];
 \filldraw (0.6,0) circle [radius=1pt];
 \filldraw (0.3,-0.7) circle [radius=1pt];
 \filldraw (0.6,-0.7) circle [radius=1pt];
\end{tikzpicture}
\end{matrix}
\begin{matrix}
\ \rightarrow \
\end{matrix}\
\begin{matrix}
\alpha \, 
\end{matrix}\
\begin{matrix}
\begin{tikzpicture}
\draw[color=red] (0,0) -- (0.9,0);
\draw[color=red] (0,0.7) -- (0.9,0.7);
\draw (0.6,0) .. controls +(0,0.2) and +(0,-0.2) .. (0.3,0.7);
 \filldraw (0.3,0) circle [radius=1pt];
 \filldraw (0.6,0) circle [radius=1pt];
 \filldraw (0.3,0.7) circle [radius=1pt];
 \filldraw (0.6,0.7) circle [radius=1pt];
\end{tikzpicture}
\end{matrix}
\begin{matrix}
 \ + \, \beta \,
\end{matrix}\
\begin{matrix}
\begin{tikzpicture}
\draw[color=red] (0,0) -- (0.9,0);
\draw[color=red] (0,0.7) -- (0.9,0.7);
\draw (0.3,0) .. controls +(0,0.2) and +(0,-0.2) .. (0.6,0.7);
 \filldraw (0.3,0) circle [radius=1pt];
 \filldraw (0.6,0) circle [radius=1pt];
 \filldraw (0.3,0.7) circle [radius=1pt];
 \filldraw (0.6,0.7) circle [radius=1pt];
\end{tikzpicture}
\end{matrix}
\end{equation}
and:
\begin{equation}
\begin{matrix}
\begin{tikzpicture}
\draw[color=red] (0,0) -- (0.9,0);
\draw[color=red] (0,-0.7) -- (0.9,-0.7);
\braid[number of strands=2,width=0.3cm,height=0.7cm,border height=0cm] a_1^{-1};
 \filldraw (0.3,0) circle [radius=1pt];
 \filldraw (0.6,0) circle [radius=1pt];
 \filldraw (0.3,-0.7) circle [radius=1pt];
 \filldraw (0.6,-0.7) circle [radius=1pt];
\end{tikzpicture}
\end{matrix}
\begin{matrix}
\ \rightarrow \
\end{matrix}
\begin{matrix}
\alpha^\ast \,
\end{matrix}\
\begin{matrix}
\begin{tikzpicture}
\draw[color=red] (0,0) -- (0.9,0);
\draw[color=red] (0,0.7) -- (0.9,0.7);
\draw (0.3,0) .. controls +(0,0.2) and +(0,-0.2) .. (0.6,0.7);
 \filldraw (0.3,0) circle [radius=1pt];
 \filldraw (0.6,0) circle [radius=1pt];
 \filldraw (0.3,0.7) circle [radius=1pt];
 \filldraw (0.6,0.7) circle [radius=1pt];
\end{tikzpicture}
\end{matrix}
\begin{matrix}
 \ + \, \beta^\ast \, 
\end{matrix}\
\begin{matrix}
\begin{tikzpicture}
\draw[color=red] (0,0) -- (0.9,0);
\draw[color=red] (0,0.7) -- (0.9,0.7);
\draw (0.6,0) .. controls +(0,0.2) and +(0,-0.2) .. (0.3,0.7);
 \filldraw (0.3,0) circle [radius=1pt];
 \filldraw (0.6,0) circle [radius=1pt];
 \filldraw (0.3,0.7) circle [radius=1pt];
 \filldraw (0.6,0.7) circle [radius=1pt];
\end{tikzpicture}
\end{matrix}
\end{equation}
Furthermore, after all the links are resolved by applying the above rule, the result will be a linear combination of planar diagrams with strands that do not cross each other. These diagrams are further decomposed into linear combinations of single-strand diagrams. For example:
\begin{equation}
\begin{matrix}
\begin{tikzpicture}
\draw[color=red] (0,0) -- (1.2,0);
\draw[color=red] (0,0.7) -- (1.2,0.7);
 \draw (0.3,0) -- (0.3,0.7);
 \draw (0.6,0) .. controls +(0,0.2) and +(0,-0.2) .. (0.9,0.7);
 \filldraw (0.3,0) circle [radius=1pt];
 \filldraw (0.6,0) circle [radius=1pt];
 \filldraw (0.9,0) circle [radius=1pt];
 \filldraw (0.6,0) circle [radius=1pt];
 \filldraw (0.3,0.7) circle [radius=1pt];
 \filldraw (0.6,0.7) circle [radius=1pt];
 \filldraw (0.9,0.7) circle [radius=1pt];
\end{tikzpicture}
\end{matrix}
\begin{matrix}
\ \rightarrow \
\end{matrix}\
\begin{matrix}
\begin{tikzpicture}
\draw[color=red] (0,0) -- (1.2,0);
\draw[color=red] (0,0.7) -- (1.2,0.7);
 \draw (0.3,0) -- (0.3,0.7);
 \filldraw (0.3,0) circle [radius=1pt];
 \filldraw (0.6,0) circle [radius=1pt];
 \filldraw (0.9,0) circle [radius=1pt];
 \filldraw (0.6,0) circle [radius=1pt];
 \filldraw (0.3,0.7) circle [radius=1pt];
 \filldraw (0.6,0.7) circle [radius=1pt];
 \filldraw (0.9,0.7) circle [radius=1pt];
\end{tikzpicture}
\end{matrix}
\begin{matrix}
 \ + \
\end{matrix}\
\begin{matrix}
\begin{tikzpicture}
\draw[color=red] (0,0) -- (1.2,0);
\draw[color=red] (0,0.7) -- (1.2,0.7);
 \draw (0.6,0) .. controls +(0,0.2) and +(0,-0.2) .. (0.9,0.7);
 \filldraw (0.3,0) circle [radius=1pt];
 \filldraw (0.6,0) circle [radius=1pt];
 \filldraw (0.9,0) circle [radius=1pt];
 \filldraw (0.6,0) circle [radius=1pt];
 \filldraw (0.3,0.7) circle [radius=1pt];
 \filldraw (0.6,0.7) circle [radius=1pt];
 \filldraw (0.9,0.7) circle [radius=1pt];
\end{tikzpicture}
\end{matrix}
\end{equation}
It is a simple exercise to verify that:
\begin{equation} 
\tau_{j,j+1}=\rho(\beta_{j,j+1}), \quad \tau^\ast_{j,j+1} = \rho(\beta^{-1}_{j,j+1}), \quad j=1,\ldots,N-1,
\end{equation}
satisfy the relations \eqref{Eq:BRelations}. For example, the generators of $\Bb_3$ become:
\begin{equation}\label{Eq:T12}
\begin{matrix}
\tau_{12} = \
\end{matrix}
\begin{matrix}
\alpha \, 
\end{matrix}\
\begin{matrix}
\begin{tikzpicture}
\draw[color=red] (0,0) rectangle (1.2,0.7);
\draw (0.6,0) .. controls +(0,0.2) and +(0,-0.2) .. (0.3,0.7);
 \filldraw (0.3,0) circle [radius=1pt];
 \filldraw (0.6,0) circle [radius=1pt];
 \filldraw (0.9,0) circle [radius=1pt]; 
 \filldraw (0.3,0.7) circle [radius=1pt];
 \filldraw (0.6,0.7) circle [radius=1pt];
 \filldraw (0.9,0.7) circle [radius=1pt];
\end{tikzpicture}
\end{matrix}
\begin{matrix}
 \ + \, \beta \,
\end{matrix}\
\begin{matrix}
\begin{tikzpicture}
\draw[color=red] (0,0) rectangle (1.2,0.7);
\draw (0.3,0) .. controls +(0,0.2) and +(0,-0.2) .. (0.6,0.7);
 \filldraw (0.3,0) circle [radius=1pt];
 \filldraw (0.6,0) circle [radius=1pt];
 \filldraw (0.9,0) circle [radius=1pt];
 \filldraw (0.3,0.7) circle [radius=1pt];
 \filldraw (0.6,0.7) circle [radius=1pt];
 \filldraw (0.9,0.7) circle [radius=1pt];
\end{tikzpicture}
\end{matrix}
 \begin{matrix}
 \ + \
 \end{matrix}
 \begin{matrix}
\begin{tikzpicture}
\draw[color=red] (0,0) rectangle (1.2,0.7);
\draw (0.9,0) -- (0.9,0.7);
 \filldraw (0.3,0) circle [radius=1pt];
 \filldraw (0.6,0) circle [radius=1pt];
 \filldraw (0.9,0) circle [radius=1pt];
 \filldraw (0.3,0.7) circle [radius=1pt];
 \filldraw (0.6,0.7) circle [radius=1pt];
 \filldraw (0.9,0.7) circle [radius=1pt];
\end{tikzpicture}
\end{matrix}
\end{equation}
and:
\begin{equation}\label{Eq:T23}
\begin{matrix}
\tau_{23} = \
\end{matrix}\
\begin{matrix}
\alpha \, 
\end{matrix}\
\begin{matrix}
\begin{tikzpicture}
\draw[color=red] (0,0) rectangle (1.2,0.7);
\draw (0.9,0) .. controls +(0,0.2) and +(0,-0.2) .. (0.6,0.7);
 \filldraw (0.3,0) circle [radius=1pt];
 \filldraw (0.6,0) circle [radius=1pt];
  \filldraw (0.9,0) circle [radius=1pt];
 \filldraw (0.3,0.7) circle [radius=1pt];
  \filldraw (0.6,0.7) circle [radius=1pt];
  \filldraw (0.9,0.7) circle [radius=1pt];
\end{tikzpicture}
\end{matrix}
\begin{matrix}
 \ + \, \beta \,
\end{matrix}\
\begin{matrix}
\begin{tikzpicture}
\draw[color=red] (0,0) rectangle (1.2,0.7);
\draw (0.6,0) .. controls +(0,0.2) and +(0,-0.2) .. (0.9,0.7);
 \filldraw (0.3,0) circle [radius=1pt];
 \filldraw (0.6,0) circle [radius=1pt];
 \filldraw (0.9,0) circle [radius=1pt];
 \filldraw (0.3,0.7) circle [radius=1pt];
 \filldraw (0.6,0.7) circle [radius=1pt];
 \filldraw (0.9,0.7) circle [radius=1pt];
\end{tikzpicture}
\end{matrix}
 \begin{matrix}
 \ + \
 \end{matrix}
 \begin{matrix}
\begin{tikzpicture}
\draw[color=red] (0,0) rectangle (1.2,0.7);
\draw (0.3,0) -- (0.3,0.7);
 \filldraw (0.3,0) circle [radius=1pt];
 \filldraw (0.6,0) circle [radius=1pt];
 \filldraw (0.9,0) circle [radius=1pt];
 \filldraw (0.3,0.7) circle [radius=1pt];
 \filldraw (0.6,0.7) circle [radius=1pt];
 \filldraw (0.9,0.7) circle [radius=1pt];
\end{tikzpicture}
\end{matrix}
\end{equation}
and, after applying the rules of calculus, we obtain:
\begin{equation}\label{Eq:T12T23T12}
\begin{matrix}
\tau_{12} \tau_{23} \tau_{12} = \
\end{matrix}
\begin{matrix}
\alpha^2 \, 
\end{matrix}\
\begin{matrix}
\begin{tikzpicture}
\draw[color=red] (0,0) rectangle (1.2,0.7);
\draw (0.9,0) .. controls +(0,0.2) and +(0,-0.2) .. (0.3,0.7);
 \filldraw (0.3,0) circle [radius=1pt];
 \filldraw (0.6,0) circle [radius=1pt];
 \filldraw (0.9,0) circle [radius=1pt]; 
 \filldraw (0.3,0.7) circle [radius=1pt];
 \filldraw (0.6,0.7) circle [radius=1pt];
 \filldraw (0.9,0.7) circle [radius=1pt];
\end{tikzpicture}
\end{matrix}
\begin{matrix}
 \ + \, \alpha\beta \,
\end{matrix}\
\begin{matrix}
\begin{tikzpicture}
\draw[color=red] (0,0) rectangle (1.2,0.7);
\draw (0.6,0) -- (0.6,0.7);
 \filldraw (0.3,0) circle [radius=1pt];
 \filldraw (0.6,0) circle [radius=1pt];
 \filldraw (0.9,0) circle [radius=1pt];
 \filldraw (0.3,0.7) circle [radius=1pt];
 \filldraw (0.6,0.7) circle [radius=1pt];
 \filldraw (0.9,0.7) circle [radius=1pt];
\end{tikzpicture}
\end{matrix}
 \begin{matrix}
 \ + \beta^2 \
 \end{matrix}
 \begin{matrix}
\begin{tikzpicture}
\draw[color=red] (0,0) rectangle (1.2,0.7);
\draw (0.9,0.7) .. controls +(0,-0.2) and +(0,0.2) .. (0.3,0);
 \filldraw (0.3,0) circle [radius=1pt];
 \filldraw (0.6,0) circle [radius=1pt];
 \filldraw (0.9,0) circle [radius=1pt];
 \filldraw (0.3,0.7) circle [radius=1pt];
 \filldraw (0.6,0.7) circle [radius=1pt];
 \filldraw (0.9,0.7) circle [radius=1pt];
\end{tikzpicture}
\end{matrix}
\end{equation}
and same for $\tau_{23} \tau_{12} \tau_{23}$.

\vspace{0.2cm}

We now construct the GNS-representation \cite{DixBook} of $T^{(1)}_N$ induced by $\Tt$. For this, one considers the linear space $\Vv_N$ that coincides with the algebra $T^{(1)}_N$ when the latter is considered only with the additive structure. To distinguish between $T^{(1)}_N$ and $\Vv_N$, we will write the elements of the latter as:
\begin{equation}\label{Eq:Vector}
|\psi\rangle = \sum_j a_j \left | \boxed{D_j} \right \rangle, \quad a_j \in \CM,
\end{equation}
Where the sum is over all available single-strand diagrams. Alternatively, $\Vv_N$ can be thought as the $\CM$-linear span of all single-strand diagrams. As such, ${\rm dim}(\Vv_N) = N^2$. Furthermore, $\Vv_N$ can be equipped with a non-degenerate scalar product, which is the unique sesqui-linear map acting on the single strand diagrams as:
\begin{equation}
\left \langle \boxed{D}, \boxed{D'} \right \rangle = \Tt\left ( \boxed{D}^\ast \cdot \boxed{D'} \right ).
\end{equation}
Then $\big (\Vv_N, \, \langle , \rangle \big )$ becomes a finite-dimensional Hilbert space.

The regular representation $\eta$ of $T^{(1)}_N$ on $\Vv_N$ is supplied by the natural action $\eta(a)|\psi\rangle = |a\psi \rangle$. Explicitly, if $\psi$ is as in \eqref{Eq:Vector}, then:
\begin{equation}\label{Eq:TnAction}
\eta \Big ( \sum_i a'_i \, \boxed{D_i}\Big )|\psi \rangle = \sum_{i,j} a'_i a_j \left |\boxed{\begin{matrix} D_i \\ D_j \end{matrix}} \right \rangle,
\end{equation}
where on the right we have the composition of single-strand diagrams. 

\vspace{0.2cm}

Let us pause and appreciate that, through the composition:
\begin{equation}\label{Eq:RegularRep}
\eta \circ \rho : \Bb_N \rightarrow {\rm GL}(\Vv_N) \simeq {\rm GL}(\CM^{N^2})
\end{equation}
we constructed a finite-dimensional representation of the braid group. Furthermore, it is immediate to check that the representation is unitary, that is, the map in \eqref{Eq:RegularRep} lands in ${\rm U}(\CM^{N^2})$ rather than the whole ${\rm GL}(\CM^{N^2})$. Let us point out that, if one tries the same procedure directly on the link diagrams, one will find that the resulting linear space is infinitely dimensional, because there are an infinite number of inequivalent link diagrams. The representation \eqref{Eq:RegularRep} is not irreducible but the irreducible representations can be easily generated once the left-ideal structure of $T^{(1)}_N$ is mapped out, which is our next task.

\vspace{0.2cm} 

The left-ideals enter the discussion in the following way. If $J \subset T^{(1)}_N$ is a left-ideal, that is, $a b \in J$ for all $a\in T^{(1)}_N$ and $b \in J$, then the linear sub-space $\Vv_J \subset \Vv_N$ spanned by all $|\psi\rangle $'s from $J$ is invariant to the regular action of $T^{(1)}_N$: 
\begin{equation}
\eta(a)|\psi \rangle = |a \psi \rangle \in \Vv_J.
\end{equation} 
Since the action of $T^{(1)}_N$ gets trapped by these sub-spaces, we can study them one at a time, instead of dealing with the large $\Vv_N$. One can quickly convince himself that the left-ideals are all direct sums of the normal left-ideals generated by one single-strand diagram. Hence an irreducible invariant sub-space of $\Vv_N$ consists of the $\CM$-linear span of single-strand diagrams originating from the same marked point at the base. There are $N$ such diagrams, hence the dimension of these invariant sub-spaces are $N$. For $N=3$, one of such invariant sub-space is generated by the following diagrams:
\begin{equation}
\begin{matrix}
\begin{tikzpicture}
\draw[color=red] (0,0) rectangle (1.2,0.7);
\draw (0.6,0) .. controls +(0,0.2) and +(0,-0.2) .. (0.3,0.7);
 \filldraw (0.3,0) circle [radius=1pt];
 \filldraw (0.6,0) circle [radius=1pt];
  \filldraw (0.9,0) circle [radius=1pt];
 \filldraw (0.3,0.7) circle [radius=1pt];
  \filldraw (0.6,0.7) circle [radius=1pt];
  \filldraw (0.9,0.7) circle [radius=1pt];
\end{tikzpicture}
\end{matrix}
\begin{matrix}
\ , \quad 
\end{matrix}\
\begin{matrix}
\begin{tikzpicture}
\draw[color=red] (0,0) rectangle (1.2,0.7);
\draw (0.6,0) .. controls +(0,0.2) and +(0,-0.2) .. (0.6,0.7);
 \filldraw (0.3,0) circle [radius=1pt];
 \filldraw (0.6,0) circle [radius=1pt];
 \filldraw (0.9,0) circle [radius=1pt];
 \filldraw (0.3,0.7) circle [radius=1pt];
 \filldraw (0.6,0.7) circle [radius=1pt];
 \filldraw (0.9,0.7) circle [radius=1pt];
\end{tikzpicture}
\end{matrix}
 \begin{matrix}
\ , \quad 
 \end{matrix}
 \begin{matrix}
\begin{tikzpicture}
\draw[color=red] (0,0) rectangle (1.2,0.7);
\draw (0.6,0) .. controls +(0,0.2) and +(0,-0.2) .. (0.9,0.7);
 \filldraw (0.3,0) circle [radius=1pt];
 \filldraw (0.6,0) circle [radius=1pt];
 \filldraw (0.9,0) circle [radius=1pt];
 \filldraw (0.3,0.7) circle [radius=1pt];
 \filldraw (0.6,0.7) circle [radius=1pt];
 \filldraw (0.9,0.7) circle [radius=1pt];
\end{tikzpicture}
\end{matrix}
\end{equation}
If one changes the base marked point, one obtains $N$ distinct such invariant sub-spaces and their direct sum is a linear space of dimension $N \times N = {\rm dim}(\Vv_N)$, hence the whole $\Vv_N$. Furthermore, the these invariant sub-spaces are canonically isomorphic, hence it is enough to study one of them. In fact, the position of the base point is irrelevant and we can think of this essentially unique invariant sub-space as the $\CM$-linear span of the diagrams:
\begin{equation}
\begin{matrix}
\varphi_1 = \
\end{matrix}
\begin{matrix}
\begin{tikzpicture}
\draw[color=red] (0,0) rectangle (1.2,0.7);
\draw (0.3,0.3) -- (0.3,0.7);
 \filldraw (0.3,0.7) circle [radius=1pt];
  \filldraw (0.6,0.7) circle [radius=1pt];
  \filldraw (0.9,0.7) circle [radius=1pt];
\end{tikzpicture}
\end{matrix}
\begin{matrix}
\ , \quad \varphi_2 = \
\end{matrix}
\begin{matrix}
\begin{tikzpicture}
\draw[color=red] (0,0) rectangle (1.2,0.7);
\draw (0.6,0.3) -- (0.6,0.7);
 \filldraw (0.3,0.7) circle [radius=1pt];
 \filldraw (0.6,0.7) circle [radius=1pt];
 \filldraw (0.9,0.7) circle [radius=1pt];
\end{tikzpicture}
\end{matrix}
 \begin{matrix}
\ , \quad \varphi_3 = \
\end{matrix}
 \begin{matrix}
\begin{tikzpicture}
\draw[color=red] (0,0) rectangle (1.2,0.7);
\draw (0.9,0.3) -- (0.9,0.7);
 \filldraw (0.3,0.7) circle [radius=1pt];
 \filldraw (0.6,0.7) circle [radius=1pt];
 \filldraw (0.9,0.7) circle [radius=1pt];
\end{tikzpicture}
\end{matrix}
\end{equation}
shown here for $N=3$. It is now easy to verify that the matrices supplied by the matrix elements:
\begin{equation}
\langle \varphi_n | \tau_{j,j+1} | \varphi_m \rangle, \quad j=1,\ldots,N_1, \quad m,n=1,\ldots,N,
\end{equation}
coincide with the $Q_{j,j+1}$ matrices introduced in section~\ref{Sec:MatrixPres}. For the case $N=3$, using $\tau_{j,j+1}$ from \eqref{Eq:T12} and \eqref{Eq:T12}, these matrices have been computed explicitly and confirmed to be identical to \eqref{Eq:QN3}.

\section{Implementation with Classical Meta-Materials}

In this section we briefly recall the algorithmic procedure from \cite{BarlasPRB2018} of transforming quantum lattice models into dynamical matrices governing the dynamics of small oscillations of coupled mechanical resonators. We also demonstrate here how to recover the standard Majorana representation of the braid group using derived physical observables.

\subsection{Classical Dynamical Matrices}
\label{Sec:ClassicalD} 

\vspace{0.2cm}

The representation of the braid group generated by the adiabatic displacements of the flux-tubes supplies an example of a non-abelian statistics that can be implemented with and observed in a classical meta-material. This is important because it represent a new and more straightforward route towards the long-sought demonstration of a controlled non-abelian braiding of arbitrary number of anyons in a physical system. Below, we briefly elaborate the main steps of this implementation.

\vspace{0.2cm}

 The key is the map $\rho$ supplied in  \cite{BarlasPRB2018}, running from linear operators over $\CM^2 \otimes \ell^2(\ZM^2)$ to linear operators over $\CM^4 \otimes \ell^2(\ZM^2)$. If we write the Hamiltonian \eqref{Eq:NewHamiltonian} as:
\begin{equation}
H = \sum_{\bm n,\bm n'} h_{\bm n',\bm n} \otimes |\bm n'\rangle \langle \bm n |, 
\end{equation} 
with $h_{\bm n',\bm n}$ being the $2 \times 2$ hopping matrices appearing in \eqref{Eq:NewHamiltonian}, then we generated a classical dynamical matrices $D$ governing the small oscillations of coupled mechanical resonators via:
\begin{equation}\label{Eq:D}
D = \omega_0^2 I + \rho(H) = \omega_0^2 I + \sum_{\bm n,\bm n'} \begin{pmatrix}{\rm Re}[h_{\bm n',\bm n}] & {\rm Im}[h_{\bm n',\bm n}] \\
- {\rm Im}[h_{\bm n',\bm n}] & {\rm Re}[h_{\bm n',\bm n}] \\
\end{pmatrix} \otimes |\bm n'\rangle \langle \bm n |,
\end{equation}
where the first term is chosen such that the spectrum of $D$ is contained by the positive real axis. As one can see, by doubling the degrees of freedom per lattice site, all entries of the dynamical matrix were made real and such dynamical matrices can be implemented in a laboratory with passive meta-materials, for example, using the platform of magnetically coupled spinners \cite{ApigoPRM2018}. This is elaborated at length in \cite{BarlasPRB2018}.

\vspace{0.2cm}

As pointed out in \cite{BarlasPRB2018}, except for a rigid shift by $\omega_0^2$, the resonant energy spectrum of $D$ is identical to that of $H$ but its degeneracy is doubled. Hence, the zero modes of $H$ double and become mid-gap normal modes for $D$ that oscillate with pulsation $\omega_0$. Furthermore, $D$ has an intrinsic and un-avoidable $U(1)$ symmetry given by the conjugation with ${\small \begin{pmatrix} 0 & I_2 \\ -I_2 & 0 \end{pmatrix}} \otimes I$. The spectrum of this symmetry operator consists of just two points $\pm \imath$, hence the Hilbert space of $D$ decouples into two dynamically invariant symmetry-sectors: 
\begin{equation}
\CM^4 \otimes \ell^2(\ZM^2) = \Pi_- \big (\CM^4 \otimes \ell^2(\ZM^2) \big ) \oplus \Pi_+ \big ( \CM^4 \otimes \ell^2(\ZM^2) \big )
\end{equation}
where $\Pi_\pm$ are the spectral projectors onto the $\pm \imath$ eigenvalues. As shown in \cite{BarlasPRB2018}, $\Pi_+ D \Pi_+$ defined over $\Pi_+ \big ( \CM^4 \otimes \ell^2(\ZM^2)\big )$  is unitarily equivalent to the original Hamiltonian $H$. In particular, it inherits the PH-symmetry, which in the classical setting is referenced from the mid-gap point:
\begin{equation}
 \widetilde \Theta_{\rm PH} \, \big ( \Pi_+ (D-\omega_0^2) \Pi_+ \big ) \, \widetilde \Theta_{\rm PH}^{-1} =  - \Pi_+ (D-\omega_0^2) \Pi_+ ,
\end{equation}
with $\widetilde \Theta_{\rm PH}$ as defined in \cite{BarlasPRB2018}:
\begin{equation}
\widetilde \Theta_{\rm PH} = J \, \rho\big (U^{-1}(\sigma_1 \otimes I) \big ) \, \Kk, \quad J = \begin{pmatrix} 0 & I_2 \\ I_2 & 0 \end{pmatrix} \otimes I.
\end{equation} 
The implication is that the whole braiding program is reproduced by the classical system defined by $D$.

\subsection{Derived physical observables and representations}

\vspace{0.2cm}

Even in the classical regime, there are interesting derived physical observables, whose dynamics supply representations of the braid group that derive from the one found in the previous sections, yet they are fundamentally different. In particular, we show below that, by using such derived physical observables, we can reproduce the standard Majorana representation of the braid group.

\vspace{0.2cm}

We assume that all initial loads of the lattice occur in the symmetry sector $\Pi_+$. An explicit and practical way to achieve such task has been elaborated at length in \cite{BarlasPRB2018}. Since $\Pi_+ (D-\omega_0^2) \Pi_+$ is unitarily equivalent to $H$ for all configurations of the flux-tubes, we will work with $H$ in the following, primarily because we already fixed the notation. Let $C\ell\big(\Hh_0,\langle,\rangle\big )$ be the complex Clifford algebra associated to the space $\Hh_0$ of zero modes with $N$ flux-tubes arranged in a particular configuration. Then, for any $\psi \in \Hh_0$, we have an element $\Gamma(\psi)$ in $C\ell\big(\Hh_0,\langle,\rangle\big )$, such that:
\begin{equation}
\Gamma(\psi)^\dagger \Gamma(\psi') + \Gamma(\psi')\Gamma(\psi)^\dagger = 2 \langle \psi,\psi' \rangle, \quad \Gamma(\Theta_{\rm PH}\psi) = \Gamma(\psi)^\dagger.
\end{equation}
We can make things more explicitly by considering the basis $\{\varphi_j\}_{j=\overline{1,N}}$ of $\Hh_0$ consisting of particle-hole symmetric modes localized at each of the flux tubes. Note that this is precisely the basis that supply the matrix form \eqref{Eq:Qs} of the braid transformations. Let:
\begin{equation}
\Gamma_j = \Gamma(\varphi_j), \quad \Gamma_j^\dagger = \Gamma_j, \quad j=1,\ldots N,
\end{equation}
which satisfy the canonical relations:
\begin{equation}
\Gamma_i \Gamma_j + \Gamma_j \Gamma_i = 2 \delta_{ij}.
\end{equation}
Note that the $\Gamma$'s can be canonically identified with matrices. Now, each of the resonant modes $\varphi_j$ can be loaded with arbitrary amplitude and phase, leading to an oscillatory state:
\begin{equation}
\psi(t) = e^{\imath \omega_0 t} \psi = e^{\imath \omega_0 t} \sum_{j=1}^N \alpha_j \varphi_j,
\end{equation}
where the coefficients $\alpha_j$ are complex amplitudes. Then:
\begin{equation}\label{Eq:GammaPsi}
\Gamma(\psi) = \sum_{j=1}^N \alpha_j \Gamma_j.
\end{equation}
The $\Gamma(\psi)$'s can play the role of our derived physical observables. Indeed, nothing stops us in feeding the measurement of the small oscillations state $\psi$ of the mechanical system into the matrix $\Gamma(\psi)$. Furthermore, note that the oscillatory state of the system can be fully recovered from \eqref{Eq:GammaPsi}:
\begin{equation}
\alpha_j = \tfrac{1}{\rm dim} {\rm Tr}\big ( \Gamma(\psi) \Gamma_j \big ), \quad j = 1,\ldots,N,
\end{equation}
where ${\rm dim}$ is the dimension of $\Gamma$'s. In fact, we can observe and manipulate products of the form $\Gamma(\psi_1) \ldots \Gamma(\psi_k)$, by creating $k$ identical copies of the system and by loading these copies, coherently, into the oscillatory states $\psi_1$, \ldots, $\psi_k$.

\vspace{0.2cm}

We now compute the induced braid operations on these derived physical observables, via the defining relation:
\begin{equation}\label{Eq:InduceActions}
\Gamma(U_{j,j+1} \psi) = \UM_{j,j+1} \, \Gamma(\psi)\, \UM_{j,j+1}^\dagger, \quad \forall \ \psi \in \Hh_0, \quad i=1,\dots,N-1,
\end{equation}
where $\UM$'s are sought inside the Clifford algebra, more precisely in the sub-algebra generated by $\Gamma_j$ and $\Gamma_{j+1}$. For simplicity, we will ignore the abelian phase factor $e^{-\imath \frac{\pi}{4}}$ in \eqref{Eq:AlphaBeta} and we will assume $\alpha=1$ and $\beta=-1$. Then the solutions to \eqref{Eq:InduceActions} are supplied by the following expressions:
\begin{equation}
\UM_{j,j+1} = \tfrac{1}{\sqrt{2}}(1 - \Gamma_j \Gamma_{j+1}), \quad \UM_{j,j+1} \UM_{j,j+1}^\dagger = \UM_{j,j+1}^\dagger \UM_{j,j+1} = I.
\end{equation}
Indeed, one can verify directly that:
\begin{align*}
\UM_{j,j+1} \Gamma(\varphi_j) \UM_{j,j+1}^\dagger=\UM_{j,j+1} \Gamma_j \UM_{j,j+1}^\dagger =  \Gamma_{j+1} = \Gamma(\alpha \varphi_{j+1}) = \Gamma(U_{j,j+1}\varphi_j),\\
\UM_{j,j+1} \Gamma_(\varphi_{j+1}) \UM_{j,j+1}^\dagger=\UM_{j,j+1} \Gamma_{j+1} \UM_{j,j+1}^\dagger = - \Gamma_{j} =\Gamma(\beta \varphi_j)= \Gamma(U_{j,j+1}\varphi_{j+1}).
\end{align*}
and for $k$ different from either $j$ and $j+1$:
\begin{equation}\nonumber
\UM_{j,j+1} \Gamma(\varphi_k) \UM_{j,j+1}^\dagger=\UM_{j,j+1} \Gamma_k \UM_{j,j+1}^\dagger =  \Gamma_k = \Gamma(\alpha \varphi_{k}) = \Gamma(U_{j,j+1}\varphi_k).
\end{equation}
By linearity, \eqref{Eq:InduceActions} follows. We now can see explicitly that the derived representation $\beta_{j,j+1} \rightarrow \UM_{j,j+1}$ coincides with the $SU(2)_2$ representation supplied by the braiding of Majorana fermions \cite{AliceaNatPhys2011}. 

\section{Conclusions and Outlook}

To observe in a laboratory the phenomena described in the previous section, we need to supply first the experimentalists with the $h$-coefficients in \eqref{Eq:D} and their variation during the adiabatic cycles. The experimentalists need to engineer couplings between four layers of square lattices of mechanical resonators, and to slowly vary in time these couplings. With the platform of magnetically coupled spinners introduced in \cite{ApigoPRM2018}, layering has been already achieved and variable couplings can be implemented by replacing the permanent magnets by electro-magnets whose strength can be programmed at will. While this were still in the planning, the exciting work \cite{ChenArxiv2019} appeared, showing an experimental demonstration and characterization of mechanical Majorana-like modes. In fact, our community is rapidly learning that classical topological meta-materials can be used quite effectively for information processing (see {\it e.g.} the recent work by  Fruchart et al \cite{FruchartArxiv2019}). 

\vspace{0.2cm}

For us, it will be extremely important to figure out how to generate and observe the derived physical observables at the hardware level. If this can be indeed achieved, then the quantum algorithms generated for the Majorana fermions program can be simulated and tested with classical hardware. An interesting aspect which was revealed to us during this work was the duality between the representations of the braid group generated with first and second quantizations, that is, between $U_{i,i+1}$ and $\UM_{i,i+1}$. As is well known \cite{WangBook,KauffmanBook2}, the latter can be generated with $N$-strands planar diagrams and our work shows that there is a relation between the representations generated with single-strand and $N$-strand planar diagrams. The passage from one to another can be achieved by simply passing to the Clifford algebra over the representation space of the former. We feel that it is important to understand this mechanism at the level of diagrams, because there are many other derived physical observables, some which naturally connect with $k$-strand diagrams, $1<k<N$. It is not excluded that dualities also exist for these cases. An even more intriguing question for us is what happens when, instead of Clifford algebra, we use generic parafermionic algebras to generate the derived physical observables. This will certainly be among our future investigations.

\section{Acknowledgements}

Yifei Liu and Yingkai Liu acknowledge financial support from the National Science Fund for Talent Training in the Basic Sciences (No. J1103208). Emil Prodan acknowledges financial support from the W. M. Keck Foundation. All authors thank Yeshiva University for hosting the 2018 summer program ``Geometry of Solid State Matter,'' during which the bulk of the work was completed.

\end{document}